\documentclass[epj]{webofc}

\usepackage[utf8]{inputenc}
\usepackage[varg]{txfonts}   
\usepackage{booktabs}
\usepackage{xcolor}
\definecolor{darkred}{rgb}{0.4,0.0,0.0}
\definecolor{darkgreen}{rgb}{0.0,0.4,0.0}
\definecolor{darkblue}{rgb}{0.0,0.0,0.4}
\usepackage[bookmarks,linktocpage,colorlinks,
    linkcolor = darkred,
    urlcolor  = darkblue,
    citecolor = darkgreen]{hyperref}
    
\usepackage{comment}
\wocname{EPJ Web of Conferences}
\woctitle{Lattice2017}
\newcommand{\be}{\begin{equation}}
\newcommand{\ee}{\end{equation}}

\def\msbar{\overline{\rm MS\kern-0.5pt}\kern0.5pt}

\begin{document}
%
\selectlanguage{english}
\title{The twelve-flavor  $\boldsymbol{\beta}$-function and dilaton tests of the sextet scalar}
\author{%
\firstname{Zoltan} \lastname{Fodor}\inst{1,2}\fnsep\thanks{presenter of contribution 405}\and
\firstname{Kieran} \lastname{Holland}\inst{3}\and
\firstname{Julius}  \lastname{Kuti}\inst{4}\fnsep\thanks{presenter of contribution 260} \and
\firstname{Daniel}  \lastname{Nogradi}\inst{5,6,7} \and
\firstname{Chik Him}  \lastname{Wong}\inst{1}
}
\institute{%
University of Wuppertal, Department of Physics, Wuppertal D-42097, Germany
\and
Juelich Supercomputing Center, Forschungszentrum Juelich, Juelich D-52425, Germany
\and
University of the Pacific, 3601 Pacific Ave, Stockton CA 95211, USA
\and
University of California, San Diego, 9500 Gilman Drive, La Jolla CA 92093, USA
\and
E\"{o}tv\"{o}s  University,  Institute  for  Theoretical  Physics
\and
MTA-ELTE Lendulet Lattice Gauge Theory Research Group, Budapest 1117, Hungary
\and
Universidad Autonoma, IFT UAM/CSIC and Departamento de Fisica Teorica, 28049 Madrid, Spain
}
\abstract{We discuss near-conformal gauge theories beyond the standard model (BSM) where interesting 
results on the twelve-flavor $\beta$-function of massless fermions in the
fundamental representation of the SU(3) color gauge group and dilaton tests of the light scalar 
with two massless fermions in the two-index symmetric tensor (sextet) representation can be viewed
as parts of the same BSM paradigm under investigation.
The clear trend in the decreasing size of $\beta$-functions at fixed renormalized
gauge coupling is interpreted as a 
first indicator how the conformal window (CW) is approached in correlation with emergent 
near-conformal light scalars. BSM model building close to the CW
will be influenced by differing expectations 
on the properties of the emergent light $0^{++}$ scalar either as a $\sigma$-particle of 
chiral symmetry breaking ($\chi SB$), or 
as a dilaton of scale symmetry breaking. The twelve-flavor $\beta$-function emerges as closest to the CW,
perhaps near-conformal, or perhaps with an infrared fixed point (IRFP) at some unexplored strong coupling inside the CW. 
It is premature to speculate on dilaton properties of the twelve-flavor model since the near-conformal realization 
remains an open question.  However, it is interesting and important to investigate dilaton tests of the light sextet scalar
whose $\beta$-function is closest to the CW in the symmetry breaking phase and emerges as the leading 
candidate for dilaton tests of the light scalar. 
We report results from high precision analysis of the twelve-flavor $\beta$-function~\cite{Fodor:2016zil} refuting its published 
IRFP~\cite{Cheng:2014jba,Hasenfratz:2016dou}. We present our objections to recent claims~\cite{Hasenfratz:2017mdh,Hasenfratz:2017qyr} 
for non-universal behavior of staggered fermions used in our analysis.
We also report our first analysis of dilaton tests of the light $0^{++}$ scalar in the sextet model and comment on related 
post-conference developments. The dilaton test is the main thrust of this conference contribution including presentation \#405 on the 
$n_f=12$ $\beta$-function and presentation \#260 on dilaton tests of the sextet model. They are both selected from the 
near-conformal BSM paradigm.
}
\maketitle
%
\section{BSM $\boldsymbol{\beta}$-functions close to the conformal window}
\begin{figure}[htb!]
	\centering
	\includegraphics[width=0.6\linewidth]{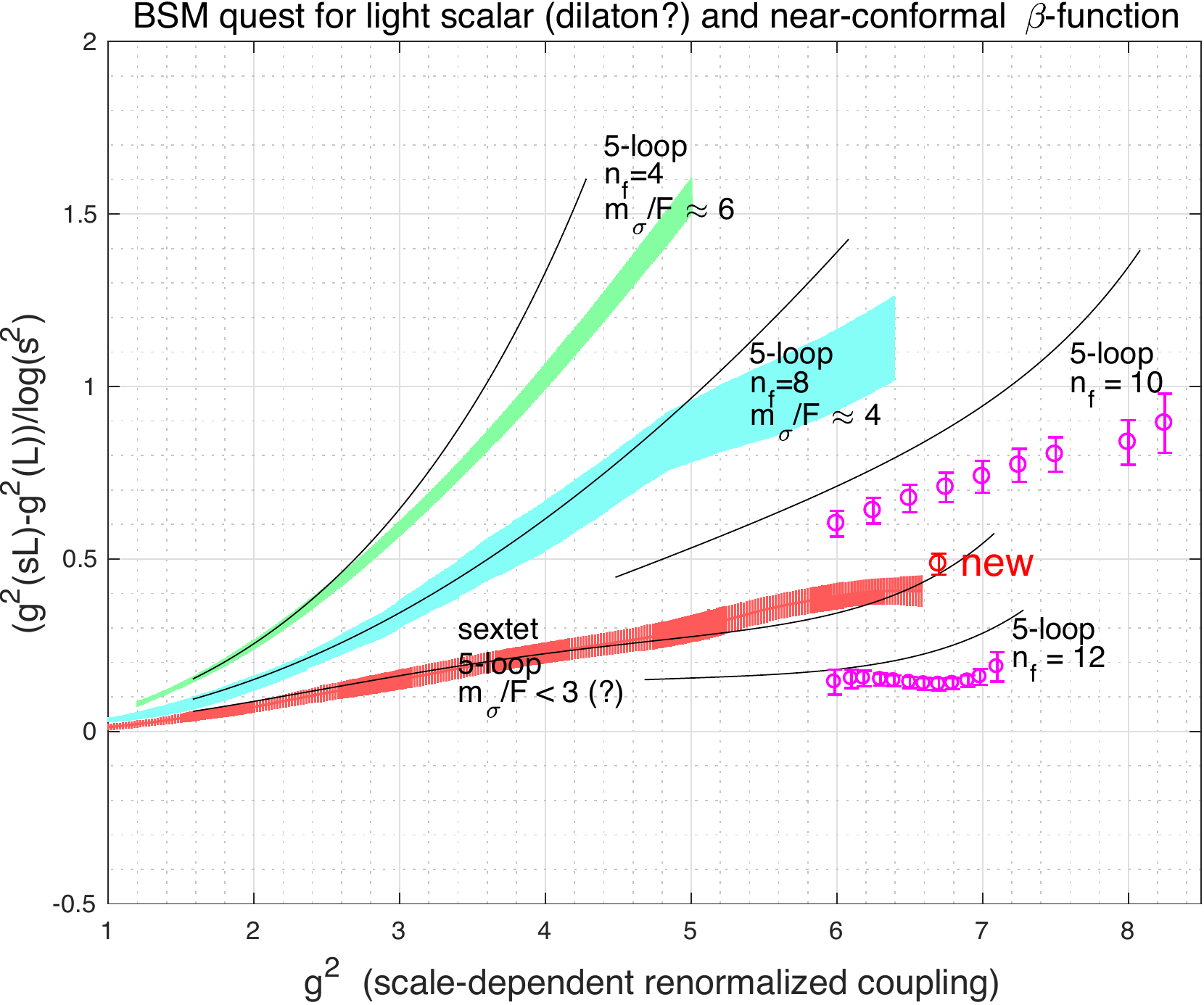}
	\caption{ \label{fig:BSMbeta}\footnotesize  The step $\beta$-functions of strongly coupled gauge theories in two 
		different fermion representations of the SU(3) gauge group are color coded. 
		The mass of the light $\sigma$-like $0^{++}$ scalar particle, as 
		a composite Higgs candidate when coupled to the Electroweak sector, is displayed in units 
		of the Goldstone decay constant $F$ in the massless fermion limit of $\chi SB$ as determined from 
		spectroscopy in each model. The striking trend of decreasing scalar mass is well established 
		as the CW is approached.
		In BSM applications $F=250~GeV$ sets the scale in physical units. 
		The $n_f=4$ $\beta$-function is from~\cite{Fodor:2012td} 
		with the $m_\sigma/F$ ratio taken from QCD,  the
		$n_f=8$ $\beta$-function is  from~\cite{Fodor:2015baa} with the $m_\sigma/F$ ratio 
		from~\cite{Aoki:2014oha,Appelquist:2016viq}, the $n_f=10$ $\beta$-function is 
		from~\cite{Fodor:2017gtj} 
		and the sextet $\beta$-function is from~\cite{Fodor:2015zna} with the $m_\sigma/F$ ratio 
		taken from~\cite{Fodor:2016pls}. The new sextet point is 
		our conference contribution~\cite{Fodor:2017die}, bridging
		the volume dependent $\beta$-function and the scale dependent $\beta$-function of the 
		p-regime in the infinite coupling limit. The (?) token next to the sextet model ratio $m_\sigma/F < 3$ 
		is an indicator that the final ratio in the chiral limit requires a second-generation analysis, underway. 
		The $n_f=12$ magenta  data points are from our \#405 conference contribution
		and from~\cite{Fodor:2017gtj}. Recent 5-loop results are not plotted for comparison with the non-perturbative results.
 	   They indicate the state of the art in the perturbative loop expansion, perhaps for future analysis. }
\end{figure}

%
The investigations of the scale-dependent renormalized gauge couplings of strongly 
coupled gauge theories and their related 
$\beta$-functions focus on near-conformal infrared behavior in a well-defined BSM paradigm.
The clear trend in the decreasing size of $\beta$-functions at fixed renormalized
gauge coupling in Fig.~\ref{fig:BSMbeta} is interpreted as a 
first indicator how the conformal window  is approached in correlation with emergent 
near-conformal light scalars. This trend in the $\beta$-function is also correlated with the slower 
rate of change in the scale-dependent gauge coupling, perhaps eventually leading to 
{\em walking} scenarios.The patterns shown in Fig.~\ref{fig:BSMbeta} were discussed
in~\cite{Fodor:2016zil} with references which are not complete here. 
BSM model building close to the CW will be influenced by differing expectations 
on the properties of the emergent light $0^{++}$ scalar as a $\sigma$-particle with chiral symmetry breaking, or 
as a dilaton of scale symmetry breaking. 
These issues were recently studied in great detail in the $n_f=8$ model with fermions
in the fundamental representation of the SU(3) color gauge group and the sextet model with two flavors in the SU(3) color gauge group
with fermions in the two-index symmetric (sextet) representation. It is tempting to ask the question, if the light scalars in the two models 
show the fingerprints of dilaton-behavior even without their explicit and difficult observations in the spectrum. Although these particles
have been exhibited directly in the simulations, less direct dilaton-tests are important to reveal their dynamical role in the 
underlying symmetry breaking pattern. We will address the issue in this proceedings contribution which includes our \#260 conference presentation.
It is interesting and important to investigate dilaton tests of the light sextet scalar
whose $\beta$-function is closest to the CW except the twelve-flavor one with undecided fate.

The $n_f=12$ model was also considered for BSM model building 
under the assumption that it is inside the CW built on an infrared fixed point which was claimed  and reported 
in~\cite{Cheng:2014jba}. 
The twelve-flavor $\beta$-function emerges as closest to the CW,
perhaps near-conformal, or perhaps with an infrared fixed point (IRFP) at some strong coupling as claimed in~\cite{Cheng:2014jba}
and later in~\cite{Hasenfratz:2016dou}. 
Although a very different conclusion was reached in~\cite{Fodor:2016zil}, contradicting the 
results of ~\cite{Cheng:2014jba,Hasenfratz:2016dou}, it is premature to speculate on dilaton properties  of the twelve-flavor model 
under near-conformal realization which remains an open question. 
For dilaton features we will focus on the sextet model.
The controversy surrounding the $n_f=12$ model 
at the conference and in some post-conference developments is briefly summarized next. 
This is an important question because the model
might show more definitive dilaton features than any of the others, if it is outside the CW.

\section{The twelve-flavor  $\boldsymbol{\beta}$-function controversy}\label{betaControversy}
\begin{figure}[thb] 
	\centering
	\includegraphics[width=0.49\linewidth,clip]{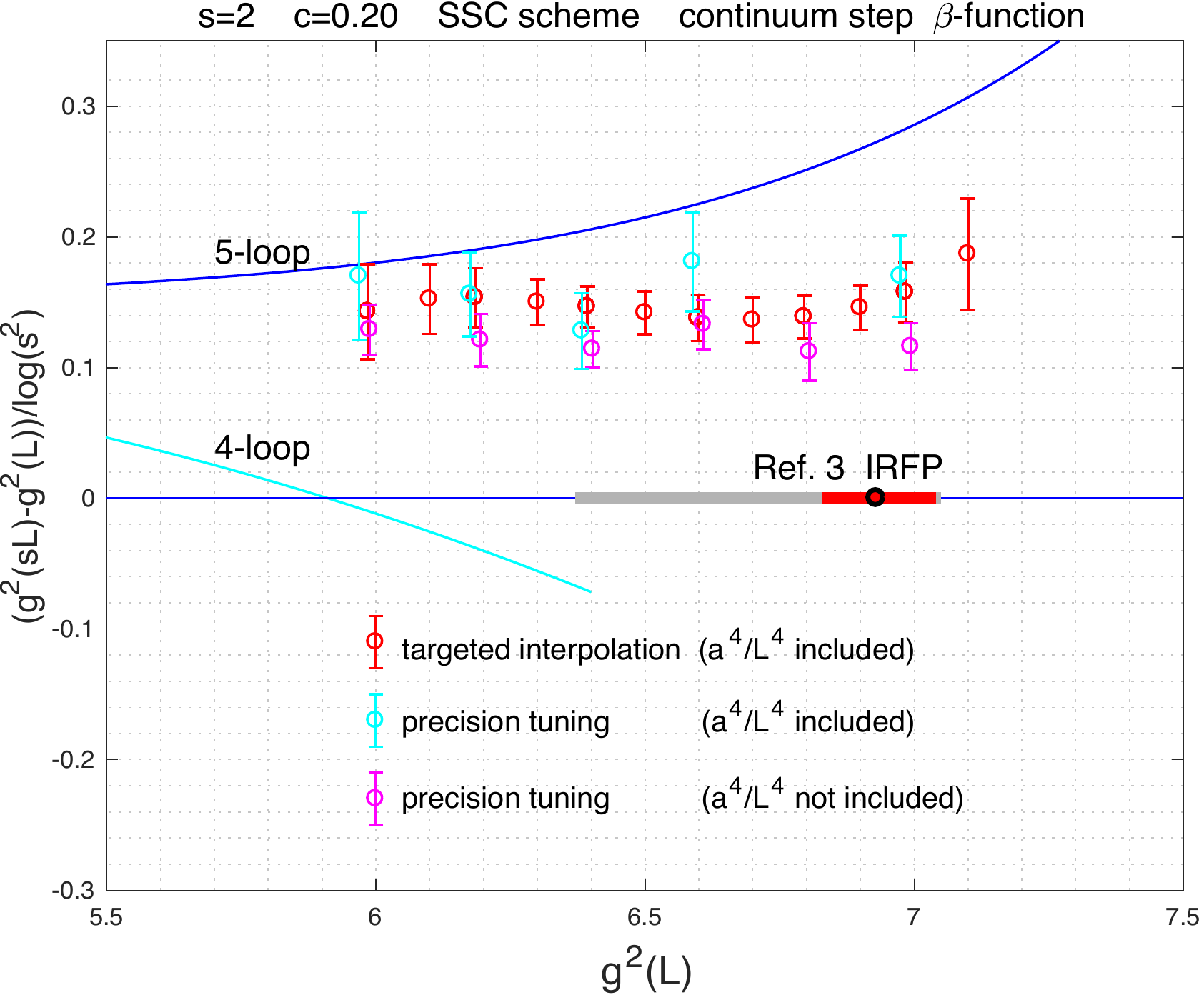}	
	\includegraphics[width=0.49\linewidth,clip]{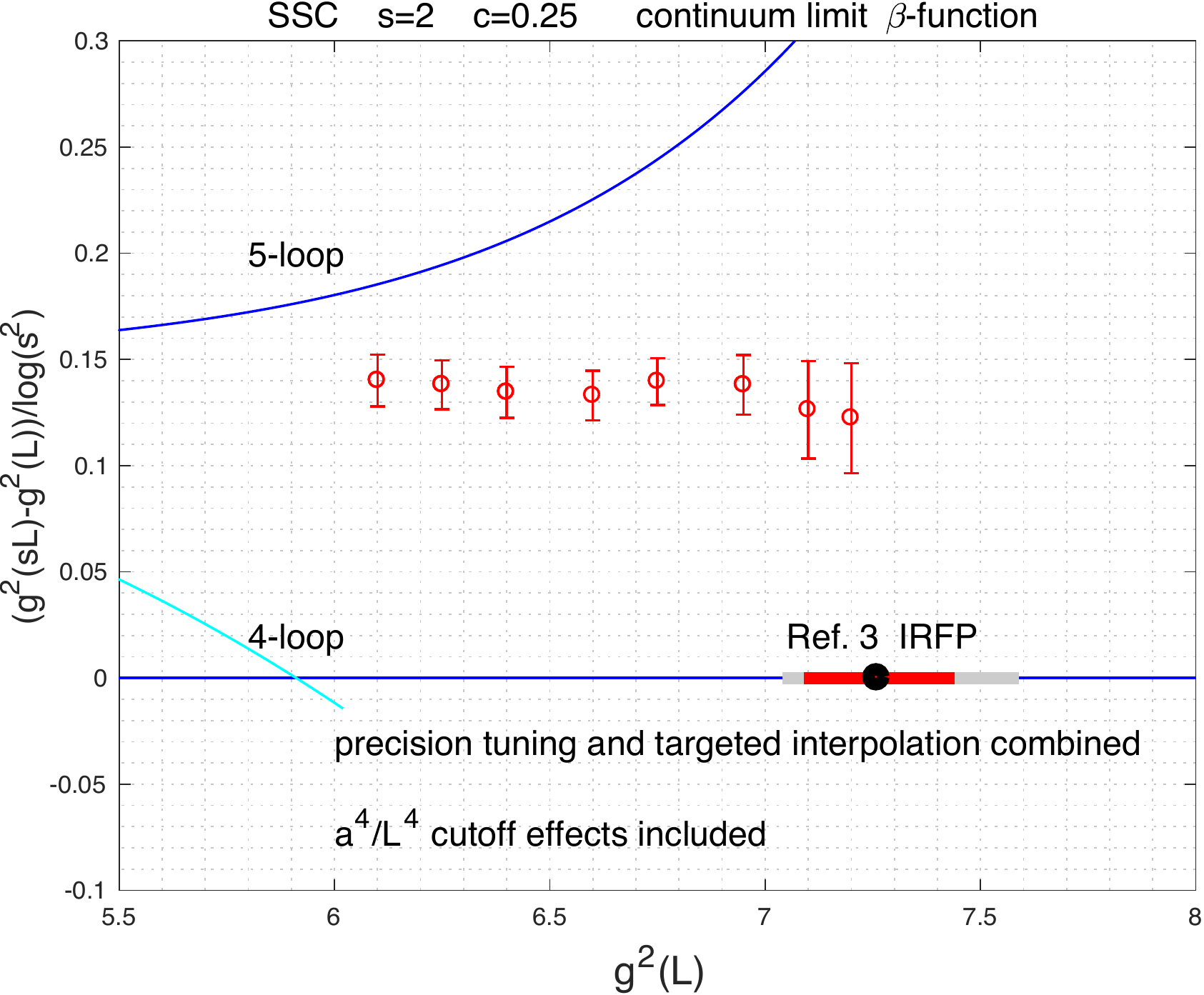}
	\caption{\footnotesize 
		Results in the figure combine~\cite{Fodor:2016zil} with the conference contribution and
		a post-conference publication~\cite{Fodor:2017gtj}.
		In the left panel results are shown for the step $\beta$-function in the $c=0.20$ 
		renormalization scheme using simple polynomial interpolation 
		for 6 combined precision tuned target runs and 3 additional auxiliary runs. Fits in extrapolation of the steps 
		to the continuum limit include the $a^4/L^4$ cutoff effects. Further details are provided for the fits in~\cite{Fodor:2017gtj}. 
		The IRFP with red bar for statistical error and grey bar for systematic estimate
		is from~\cite{Hasenfratz:2016dou}.  The controversy is self-evident.                                    
		The right panel shows results  for the SSC gradient flow  in the $c=0.25$ renormalization scheme with step size $s=2$ 
		using simple polynomial interpolation for 6 combined precision tuned target runs and 3 additional auxiliary runs. 
		The location of the IRFP in the
		$c=0.25$ scheme of~\cite{Hasenfratz:2016dou} is somewhat shifted to the right from the location of the IRFP  in the
		$c=0.20$ scheme.
		The 4-loop and recent 5-loop results for the $\beta$-function are shown 
		with similar limited purpose as explained in Fig.~\ref{fig:BSMbeta}.
	} 
	\label{fig:BSMbeta1}
\end{figure}

A conformal infrared fixed point (IRFP) of the $\beta$-function was reported earlier
with critical gauge coupling $g_*^2 \approx 6.2$ and interpreted as conformal behavior
of the much studied BSM gauge theory with twelve massless fermions
in the fundamental representation of the SU(3) color gauge group~\cite{Cheng:2014jba}. 
This result was claimed to confirm the original finding of the IRFP in~\cite{Appelquist:2007hu,Appelquist:2009ty}.
In disagreement with~\cite{Cheng:2014jba,Appelquist:2007hu,Appelquist:2009ty}, the IRFP was  
refuted in~\cite{Fodor:2016zil}. Recently, responding to the negative findings
in~\cite{Fodor:2016zil}, the authors of~\cite{Cheng:2014jba} moved the IRFP to a revised new  location 
$g_*^2 \approx 7$ in~\cite{Hasenfratz:2016dou}. This new location of the IRFP was refuted at
the conference and in a post-conference publication~\cite{Fodor:2017gtj} with all results combined in Fig.~\ref{fig:BSMbeta1}.  
The relocation of the IRFP in~\cite{Hasenfratz:2016dou} followed the announcement 
of a new IRFP  with ten massless fermion flavors in the fundamental representation of 
the SU(3) color gauge group~\cite{Chiu:2016uui,Chiu:2017kza}.
The claim in~\cite{Chiu:2016uui,Chiu:2017kza}  would imply that 
the theory with twelve flavors must also be conformal and the lower edge of the 
conformal window (CW) of multi-flavor BSM theories with fermions in the fundamental representation 
would be located below ten flavors.  No trace of the reported IRFP with ten flavors was found from
high precision simulations in large volumes~\cite{Fodor:2017gtj}.

\begin{figure}[htb!]
	\centering
	\includegraphics[width=0.75\linewidth]{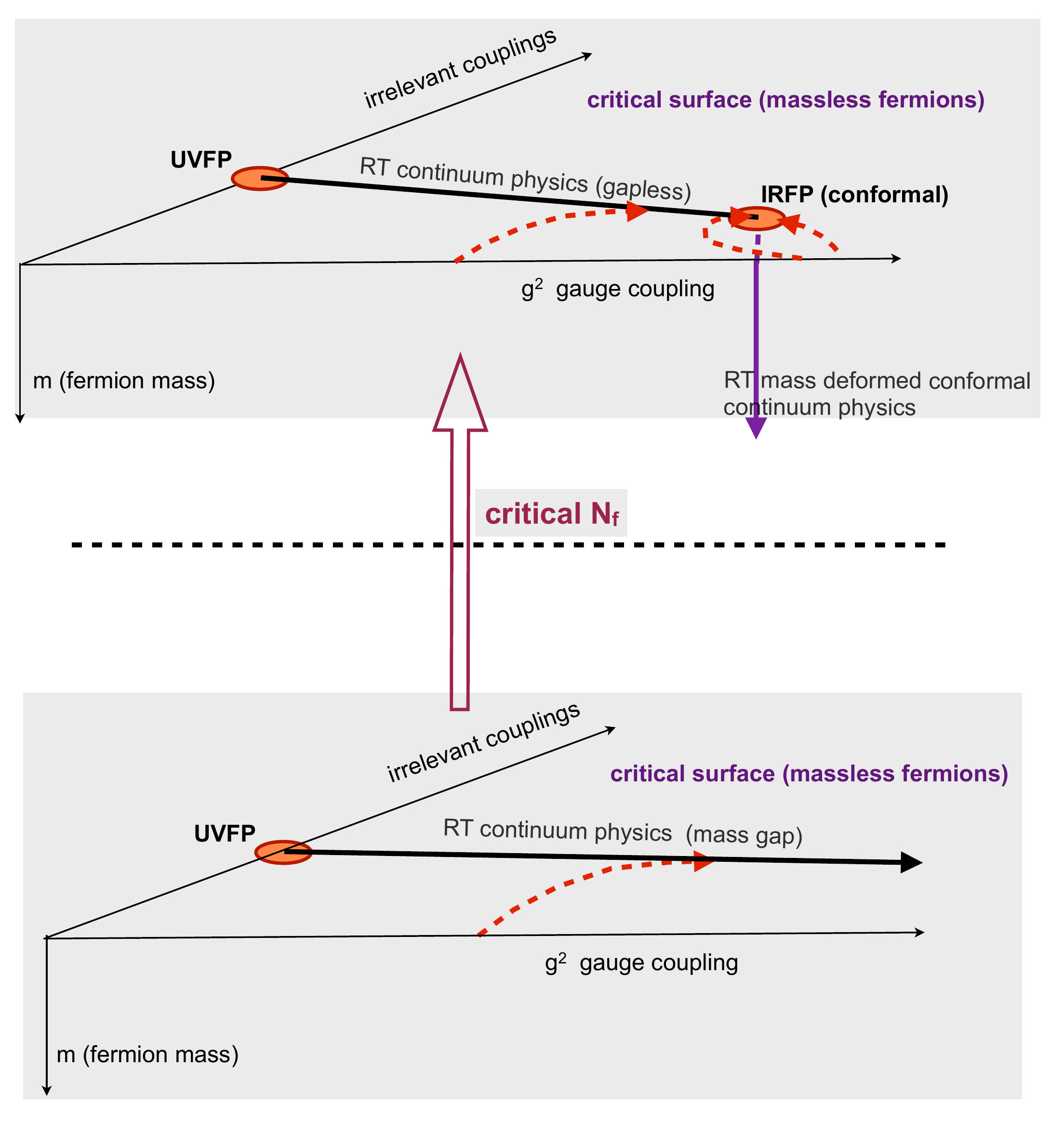}
	\caption{\label{fig:RTplot}\footnotesize  The critical surface of massless fermions is illustrated below the conformal window (lower part of the plot)
	and inside the CW (upper part). The Renormalized Trajectory (RT), built on the UV fixed point and 
	representing continuum physics is illustrated 
	in both phases. 
	The role of relevant fermion mass deformations is also shown in the plot. In QCD-like theories (lower part) the 
    universality of RT built with staggered fermions is not challenged by the authors of~\cite{Hasenfratz:2017mdh,Hasenfratz:2017qyr}.
    It is challenged inside the CW in the upper part of the plot.}
\end{figure}

At the conference, an explanation was suggested to resolve the controversy~\cite{Hasenfratz:2017mdh}.  Accordingly, the authors claim
that staggered formulations of a conformal system are not in the same universality class as continuum-like (domain wall) fermions 
unless the taste breaking terms of staggered fermions vanish at the conformal IRFP. The authors of~\cite{Hasenfratz:2017mdh} argue that the 
universality class of staggered fermions cannot be answered perturbatively and a non-perturbative calculation is needed
to compare the critical behavior of staggered and continuum-like domain wall fermions at a conformal IRFP. 
They claim that they did that because they
exhibited  $\beta$-functions for the same $n_f=12$ model in identical renormalization schemes where the  staggered $\beta$-function differs
at finite renormalized gauge coupling from the $\beta$-function of domain wall fermions. Hence no controversy, just staggered fermions
failing in conformal studies.
A post-conference publication 
reiterated and further elaborated the same argument~\cite{Hasenfratz:2017qyr}.
We see major flaws in this argument,  illustrated in Fig.~\ref{fig:RTplot} with the rebuttal below it.
\begin{enumerate}[(a)]
	
	\item At the origin of the contention is the $n_f=12$ model with the 
	authors of~\cite{Cheng:2014jba,Hasenfratz:2016dou,Hasenfratz:2017mdh} claiming an IRFP and  
	the authors of~\cite{Fodor:2016zil,Fodor:2017gtj} refuting it. Since both groups use the same staggered fermion formulation
	and the same continuum renormalization scheme, staggered fermion universality or its failure has nothing to do 
	with the disputed controversy which originated from staggered fermion results used by both groups. 
	We do not see how to avoid the logical conclusion that some staggered results must be incorrect. 
	New results from ~\cite{Hasenfratz:2016dou,Hasenfratz:2017mdh}
	using domain wall fermions is a separate issue. 

	\item 
	To understand better the controversy in using and comparing domain wall fermions with staggered fermions,
	we recall first that  4d gauge theories with isolated conformal fixed points are not known.  
	In the critical surface and inside the CW, the Renormalized Trajectory (RT) connects the UV fixed point 
	with the conformal fixed point $g*$, as illustrated by the upper part of Fig.~\ref{fig:RTplot}. Representing the continuum theory,
	at any point on the RT, outside the conformal fixed point, the Green functions are asymptotically free,
	in the conformal fixed point itself they are not, but this is irrelevant. The renormalized scale-dependent coupling
	can get arbitrary close to $g*$ but flowing into it takes infinite time. The interchange of two limits
	should not be added to the confusion. Any point on the renormalized trajectory representing continuum physics,
   built on the UV fixed point, has the continuum limit of the staggered formulation with taste symmetry restored,
   just as universal as in QCD. In this regard there should be no difference  between the two RT in the lower
	and upper parts of Fig.~\ref{fig:RTplot}. 
	
    \item The authors of~\cite{Cheng:2014jba,Hasenfratz:2016dou,Hasenfratz:2017mdh} obtain their results
	building their own analysis on the  UV fixed point when they extrapolate to the continuum limit with $a^2/L^2$ cutoff dependence. 
	They use the Renormalized Trajectory connecting the UV fixed point with the conformal fixed point, 
	if there is one in the model, as they claim.
	This has consequences. In the $\beta$-function of the $n_f=12$ model shown in Fig. 4 of~\cite{Hasenfratz:2017qyr} 
	there shouldn't  be  a different $\beta$-function with domain wall 
	fermion when compared with their staggered result, both built on the UV fixed point. 
	The two results in Fig. 4 of~\cite{Hasenfratz:2017qyr} are inconsistent.
	
	\item Since staggered fermions at $n_f=12$ are built on a UV fixed point at zero gauge coupling,
	relevant or marginal operators, like in the examples of the 3D statistical models in~\cite{Hasenfratz:2017mdh}, 
	cannot be added to the staggered lattice fermion action which has correct locality and universality properties. 
	The explicit construction is well-known in the literature. 
\end{enumerate}

\section{Dilaton tests in the sextet model}
The light  $0^{++}$ scalar, discovered in the $n_f=8$ model by the LatKMI collaboration~\cite{Aoki:2014oha} in the year 2013 and
also confirmed later by the LSD collaboration~\cite{Appelquist:2016viq}, provided strong motivations for 
$\sigma$-particle and dilaton studies by the LatKMI and LSD collaborations.
The LatHC project, concurrently with the findings of the LatKMI group, has discovered in the year 2013 an even lighter $0^{++}$ 
scalar in the sextet model~\cite{Fodor:2014pqa}. The two groups, LatKMI and LatHC, played pioneering role in the discovery 
of the the $0^{++}$  light scalar of the $n_f=8$ model and the light scalar of sextet in the year 2013 with new follow-up work 
from the LSD collaboration on the $n_f=8$ model~\cite{Appelquist:2016viq}. 
The early pioneering LatKMI and LatHC discoveries of the year 2013 were reviewed in~\cite{Kuti:2014epa}.

Soon it became clear that some improved effective theory of the intriguing coupled dynamics between the emergent low mass scalar 
($0^{++}$ $\sigma$-particle) and Goldstone pions from $\chi SB$ was needed.
This dynamics is expected to be different from what we know about the $\sigma$-particle in QCD. 
Precise Electroweak scale setting in terms of the pion decay constant ${ F_\pi}$ in the chiral limit has 
remained a very important goal. If the emergent light scalar has dilaton signatures from scale symmetry breaking 
close to the conformal window, it will change the perspective on the effective theory of infrared pion dynamics coupled 
to the $0^{++}$ state. 
Looking for dilaton signatures of the light $0^{++}$ scalar of near-conformal gauge theories  in the framework of low-energy 
effective theories, has a history going back several years. 
A pilot study was published first to investigate dilaton footprints in the $n_f=8$ model~\cite{Matsuzaki:2013eva} building the 
dilaton effective theory with ingredients from prior work, including input from~\cite{Leung:1989hw}. The early
effort of~\cite{Matsuzaki:2013eva} was followed with more comprehensive dilaton inspired tests of the $n_f=8$ model while 
looking for walking signatures of the scale-dependent gauge coupling and its $\beta$-function~\cite{Aoki:2016wnc}. 

Our collaboration started early work in the sextet model on the coupled dynamics of the light scalar and light 
pseudo-Nambu-Goldstone (pngb) pions.
It became clear that the linear $\sigma$-model without modifications would not work in the regime explored by the simulations
but generalizations introduced many new parameters hindering comprehensive analysis. From the dilaton perspective, 
influential recent papers provided new guidance for the general framework\cite{Golterman:2016lsd,Kasai:2016ifi,Golterman:2016cdd}.
Added impetus was provided from the investigation of the $n_f=8$ model by new work from Appelquist et al. testing their well-reasoned 
and simple effective theory for the dilaton~\cite{Appelquist:2017wcg},  compatible with the general framework of~\cite{Golterman:2016lsd}.
Success of the dilaton tests was reported in~\cite{Appelquist:2017wcg} confirming scaling relations of the underlying effective dilaton theory.
The authors emphasized their effort to determine the dilaton potential from lattice data as one of the key novelties of their framework. 

In this conference contribution we test the framework of~\cite{Appelquist:2017wcg} in the sextet model. While working on the tests, 
a new paper was submitted~\cite{Appelquist:2017vyy} which included some dilaton analysis of the sextet model against our reservations. 
The reservations originated from the unacceptable shortcuts in the analysis~\cite{Appelquist:2017vyy}, 
reading off old data from old plots of our earlier papers and adding ad hoc 
"systematic errors" to the so-called data without knowing the real ones, and reporting success for sextet dilaton analysis.
We do not find acknowledgment of our own work in~\cite{Appelquist:2017vyy} although our conference report and its upcoming 
publication was known to the authors.
Here we provide results from our own tests, while in the limited space we need to refer to~\cite{Appelquist:2017wcg,Appelquist:2017vyy} 
for notation and technical explanation. Keeping the same notations as the authors of~\cite{Appelquist:2017wcg,Appelquist:2017vyy} 
will make the comparisons more informative for the reader. First, we describe the input which goes into the analysis, followed by two
critically important dilaton tests.

\subsection{Finite size scaling analysis (FSS) of the input $\boldsymbol{M_\pi}$ and $\boldsymbol{F_\pi}$ data}\label{FSS}
We have our data set from a very large number of gauge ensembles at three lattice spacings with a range $m=0.0010-0.0080$ 
of fermion masses $m$ and lattices sizes from $32^3\times 64$ to $64^3\times 96$. Only the analysis of dilaton test results 
from bare gauge couplings $\beta = 6/g^2$ at $\beta=3.20$ and $\beta=3.25$ are reported here. Ensembles at the finest lattice 
spacing with $\beta=3.30$ are under ongoing investigation. 
\begin{figure}[htb!]
	\centering
	\includegraphics[width=0.49\linewidth]{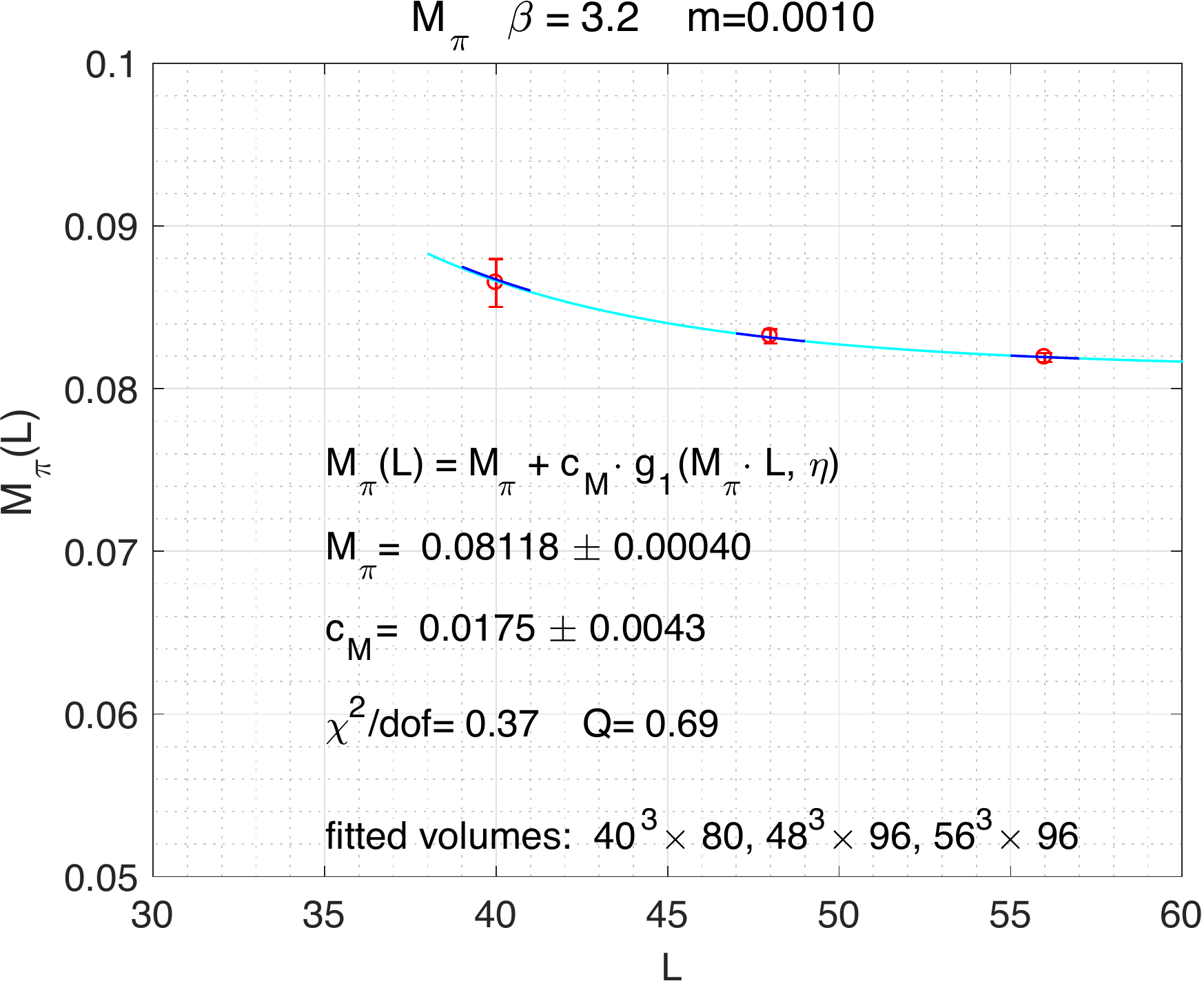}
	\includegraphics[width=0.49\linewidth]{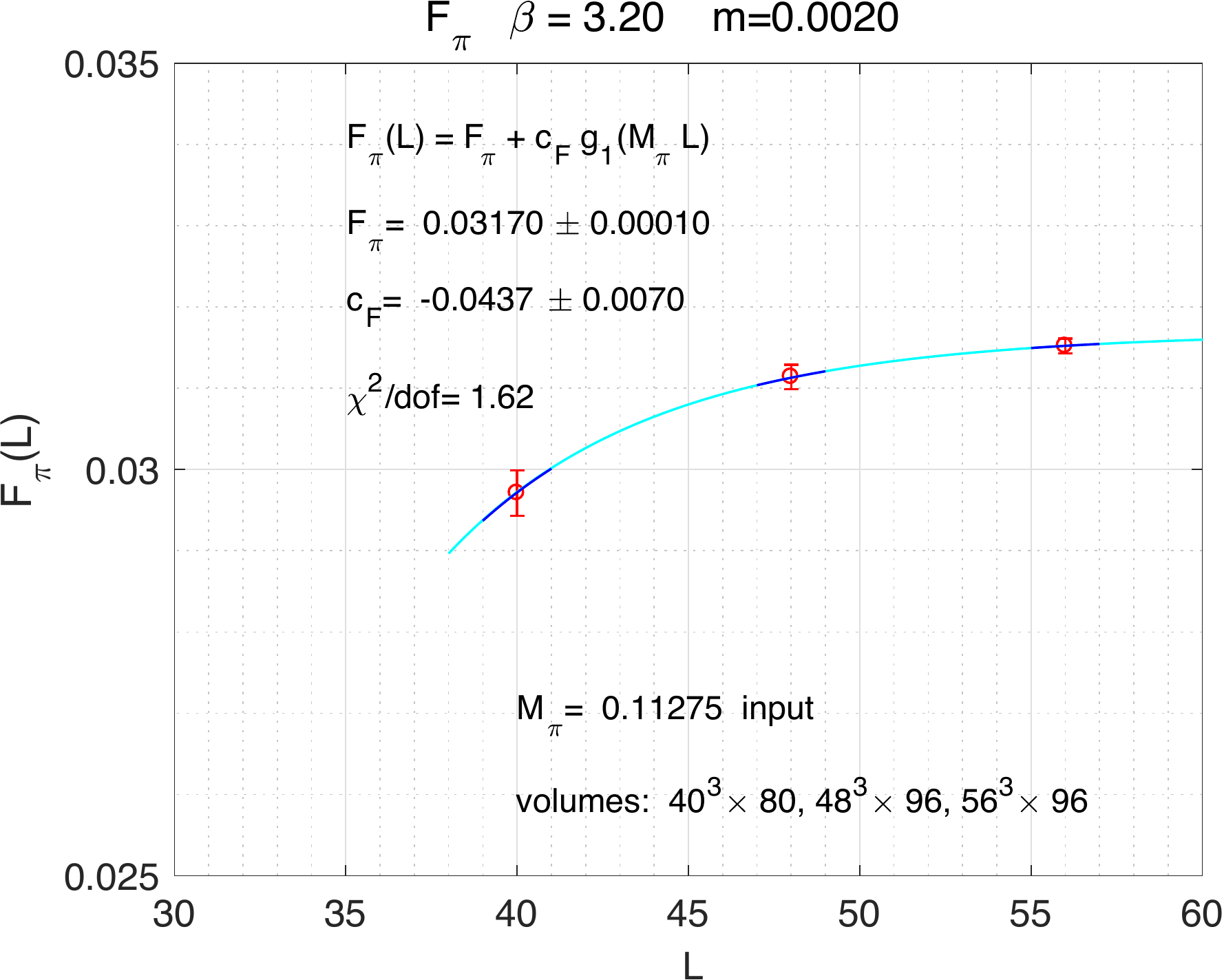}\\
	\includegraphics[width=0.49\linewidth]{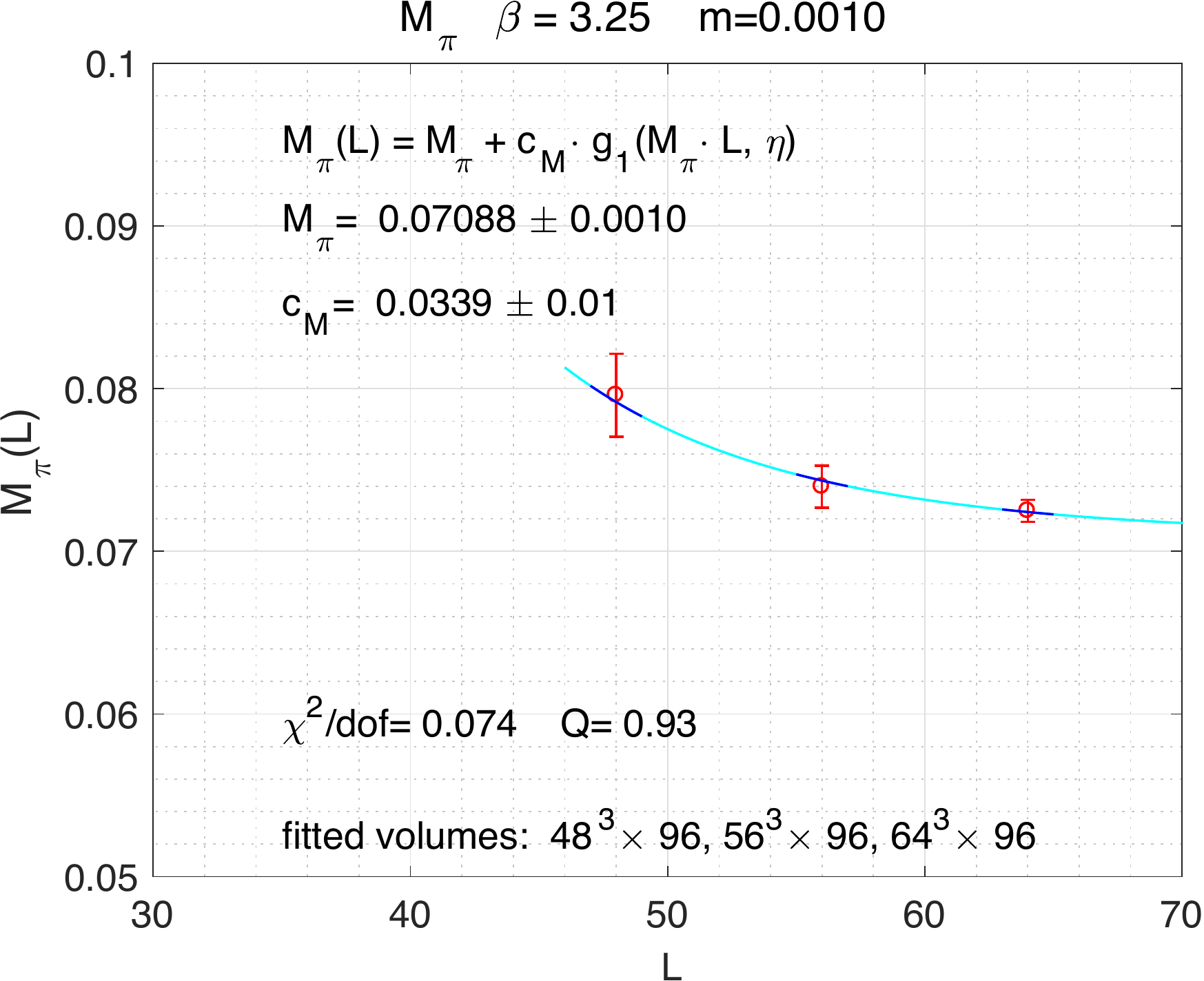}
	\includegraphics[width=0.49\linewidth]{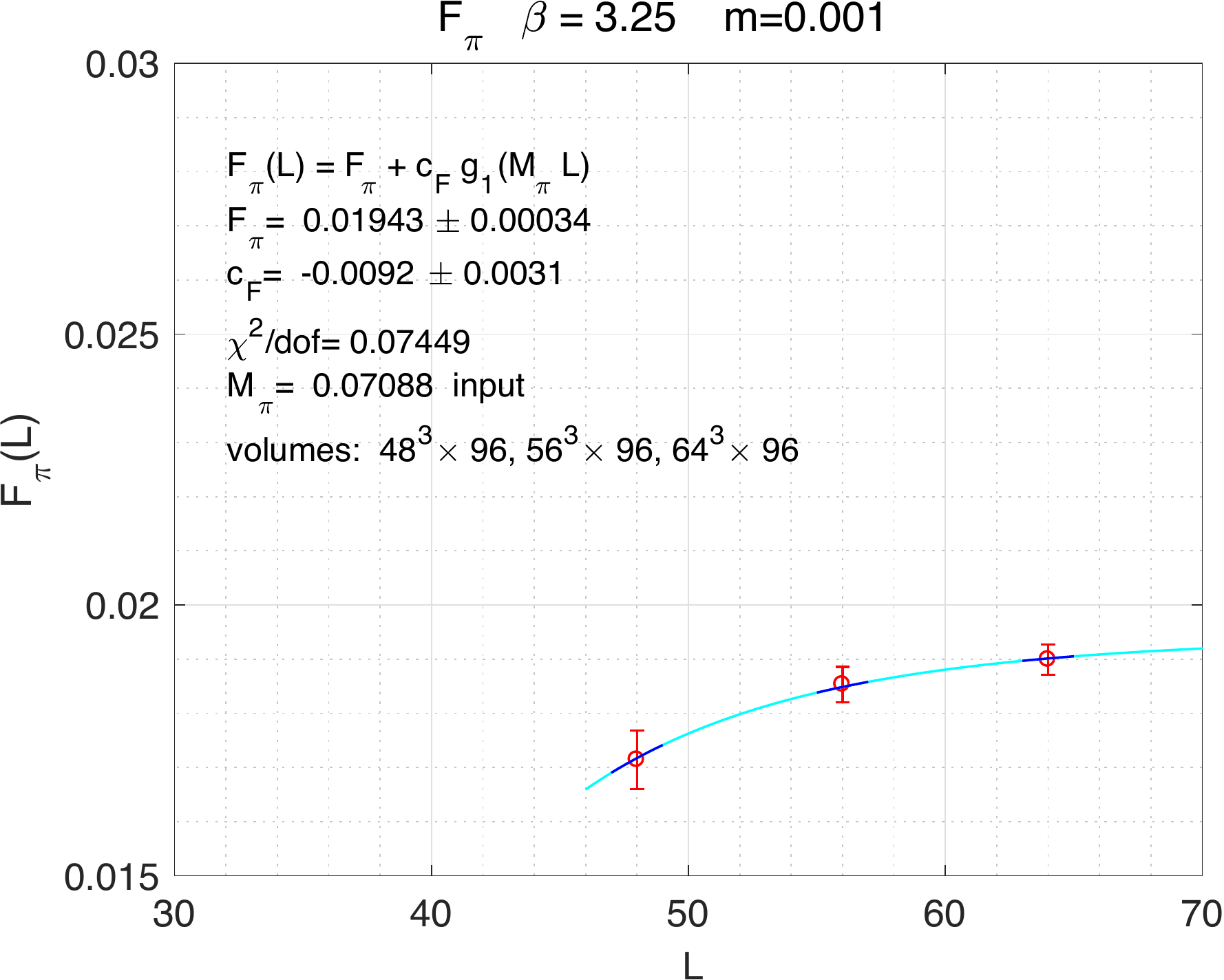}	
	\caption{\label{fig:FSSpion} Typical FSS fits, shown here at two different lattice spacings, are discussed in the main text.}
\end{figure}
For the dilaton analysis of the scaling tests proposed 
in~\cite{Appelquist:2017wcg,Appelquist:2017vyy} we use five input values of $M_\pi$ and $F_\pi$ at each of 
the two $\beta = 6/g^2$ values. The twenty input data is the outcome of FSS analysis which is necessary to avoid systematic errors in the tests.
At each of the fermion masses in the $m=0.0010,0.0015,0.0020,0.0030,0.0040$ range three lattice volumes are used so that altogether 
sixty $M_\pi$ and $F_\pi$ data were used from double-jackknife analysis of effective masses and pion decay constants using Rwall  
(random wall source) pion correlators.
We use an ansatz with an infinite sum $g_1$ of Bessel functions dependent on the aspect ratio $L_t/L_s$ of the lattice volume 
to account for Goldstone bosons wrapping around the finite volume~\cite{Gasser:1986vb}, $M_\pi(L) = M_\pi + c_M g_1(M_\pi L)$ 
and $F_\pi(L) = F_\pi + c_F g_1(M_\pi L)$, where the complicated sum $g_1$ is evaluated numerically. 
At 1-loop in chiral perturbation theory $c_M = M_\pi^2/(64 \pi^2 F_\pi^2)$, but we leave the prefactors $c_M$ and $c_F$ of the $g_1$ function 
as free parameters to be fitted. In Fig.~\ref{fig:FSSpion} we show examples of such infinite volume extrapolations 
for the Goldstone boson mass $M_\pi$ and the decay constant $F_\pi$. These figures are typical: the volume effect is 
relatively small but quite visible and is well described by the ansatz. Note that the infinite volume mass $M_\pi$ and decay 
constant $F_\pi$ are determined in simultaneous fits, from inputs of 
finite volume $M_\pi(L)$ and $F_\pi(L)$ data together with the $c_M$ and $c_F$ amplitudes at each value of the fermion mass $m$.
In the final analysis additional FSS effects, if they exist, like one from wrap-around of the light scalar itself, should be also addressed. 

\subsection{Model-independent test of the dilaton effective action}
\begin{figure}[htb!]
	\centering
	\includegraphics[width=0.49\linewidth]{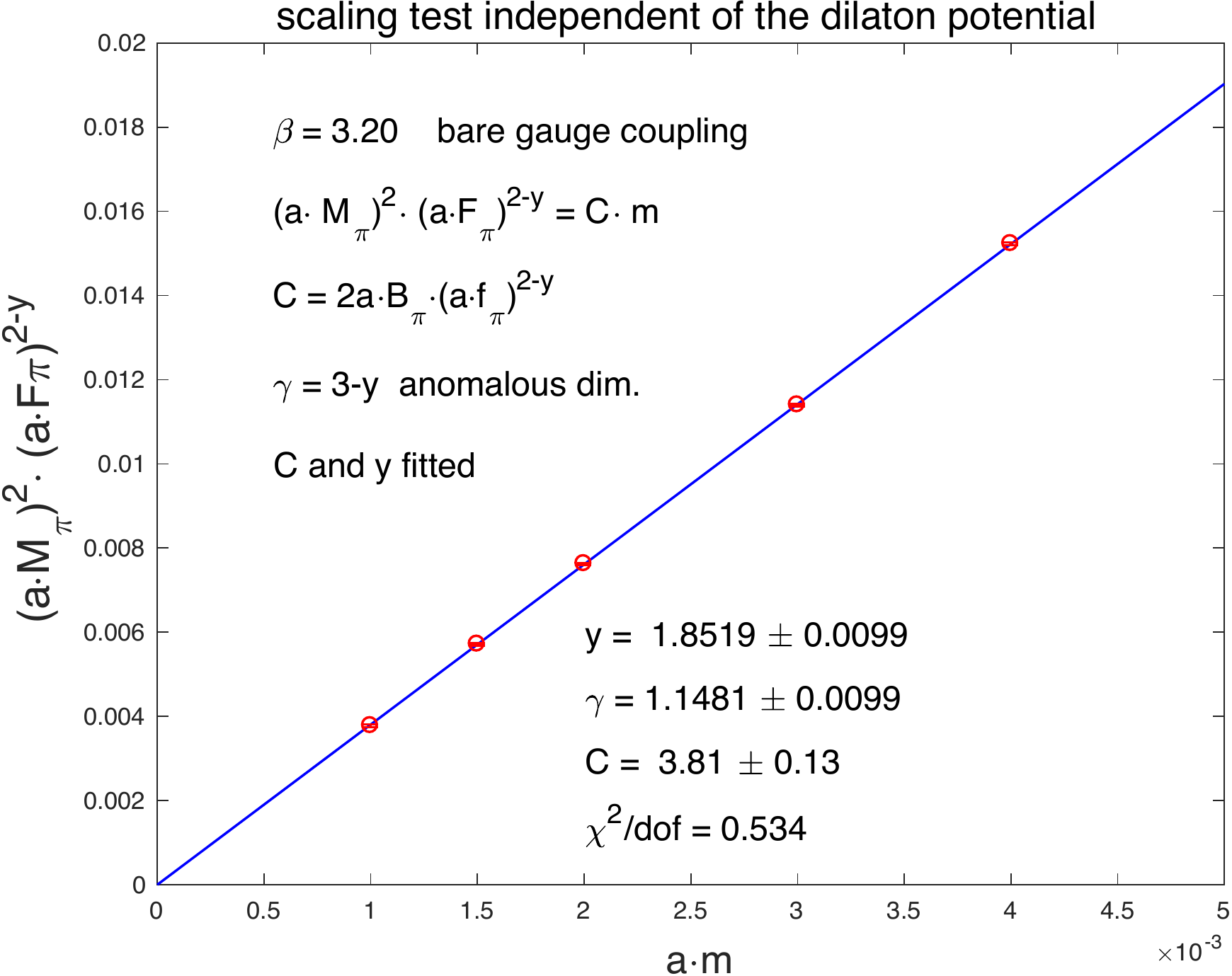}
	\includegraphics[width=0.49\linewidth]{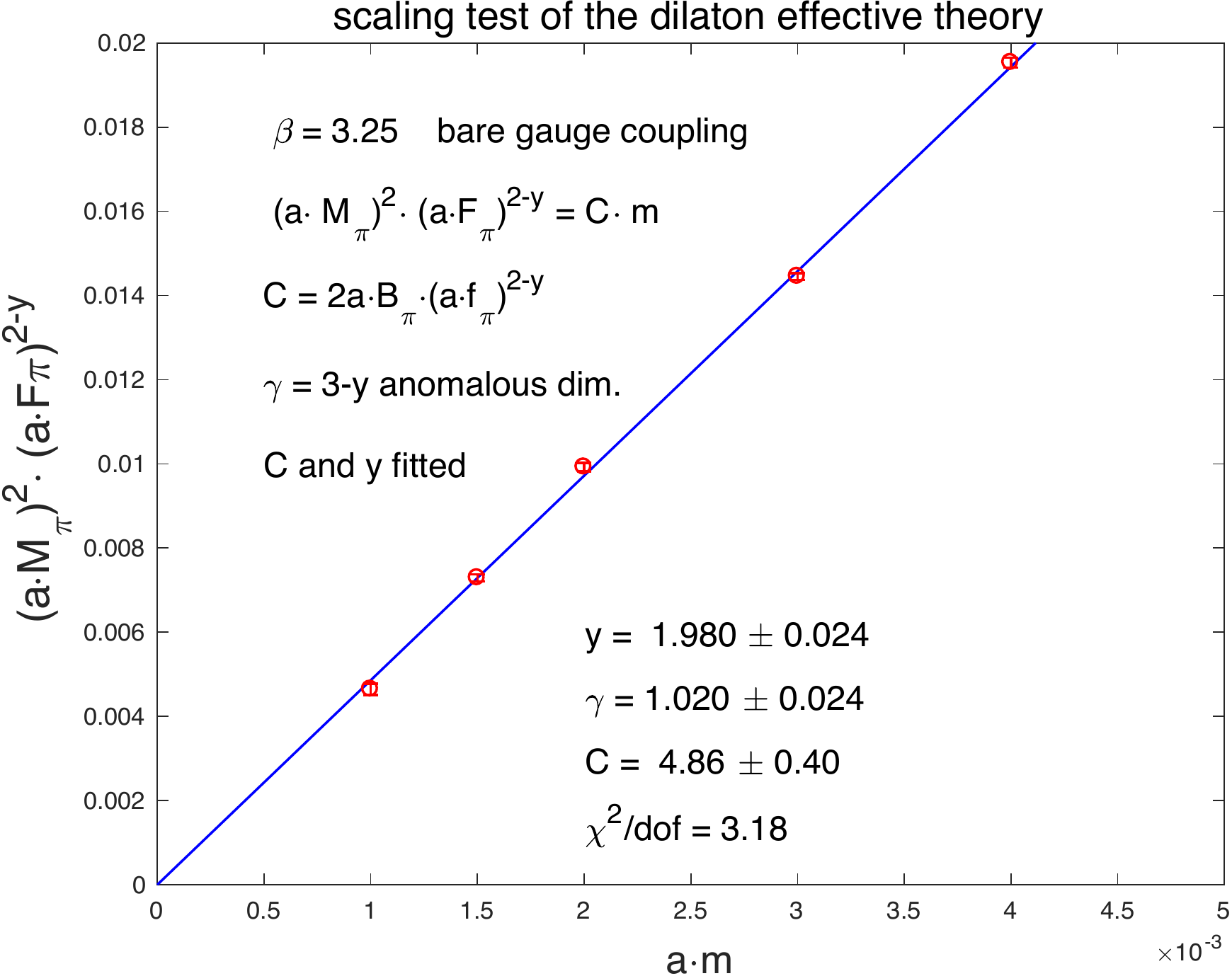}
	\caption{\label{fig:dilaton1} The model-independent test of the dilaton effective theory is shown
	at two different lattice spacings. The fitted relation of the figure is discussed in the main text.}
\end{figure}

A tree-level scaling test was proposed in~\cite{Appelquist:2017wcg,Golterman:2016lsd} which is not dependent on the
choice of the dilaton potential. The results for the sextet model are shown in Fig.~\ref{fig:dilaton1} for two 
gauge couplings.
Accordingly, a general fingerprint of the dilaton is the $M_\pi^2\cdot F_\pi^{2-y} = C\cdot m$ linear relation 
in the bare fermion mass $m$. The FSS corrected values of $M_\pi$ and $F_\pi$ are used as input in the figure.
The fitted constant $C=  2a\cdot B_\pi(af_\pi)^{2-y}$ depends on the constant $B_\pi$ of the dilaton Lagrangian
(also familiar from chiral perturbation theory), on the pion decay constant $f_\pi$ in the chiral limit, and on the 
anomalous exponent $y$. The exponent $y$ is related to the mass anomalous dimension $\gamma$ 
by the relation $y=3-\gamma$. Although $B_\pi$ itself requires renormalization, the combination $B_\pi\cdot m$
which enters into the test is RG invariant. Note the difference in notation between $F_\pi$ at finite fermion mass 
and  $f_\pi$ in the $m\rightarrow 0$ chiral limit.

The test result at $\beta=3.20$ is nearly perfect but at $\beta=3.25$ the larger $\chi^2$ value is somewhat problematic.
We investigated this in great detail and it remains unclear if the $\beta=3.25$ fit will improve in more refined analysis.
The statistical analysis is robust and checked in several ways. It is somewhat unusual to fit the expression 
$M_\pi^2\cdot F_\pi^{2-y} = C\cdot m$ where data input depends on the fitted parameter $y$. It is easy to see that this is 
still a Maximum Likelihood (ML) procedure under certain conditions which can be tested. 
\begin{figure}[htb!]
	\centering
	\includegraphics[width=2.55 in]{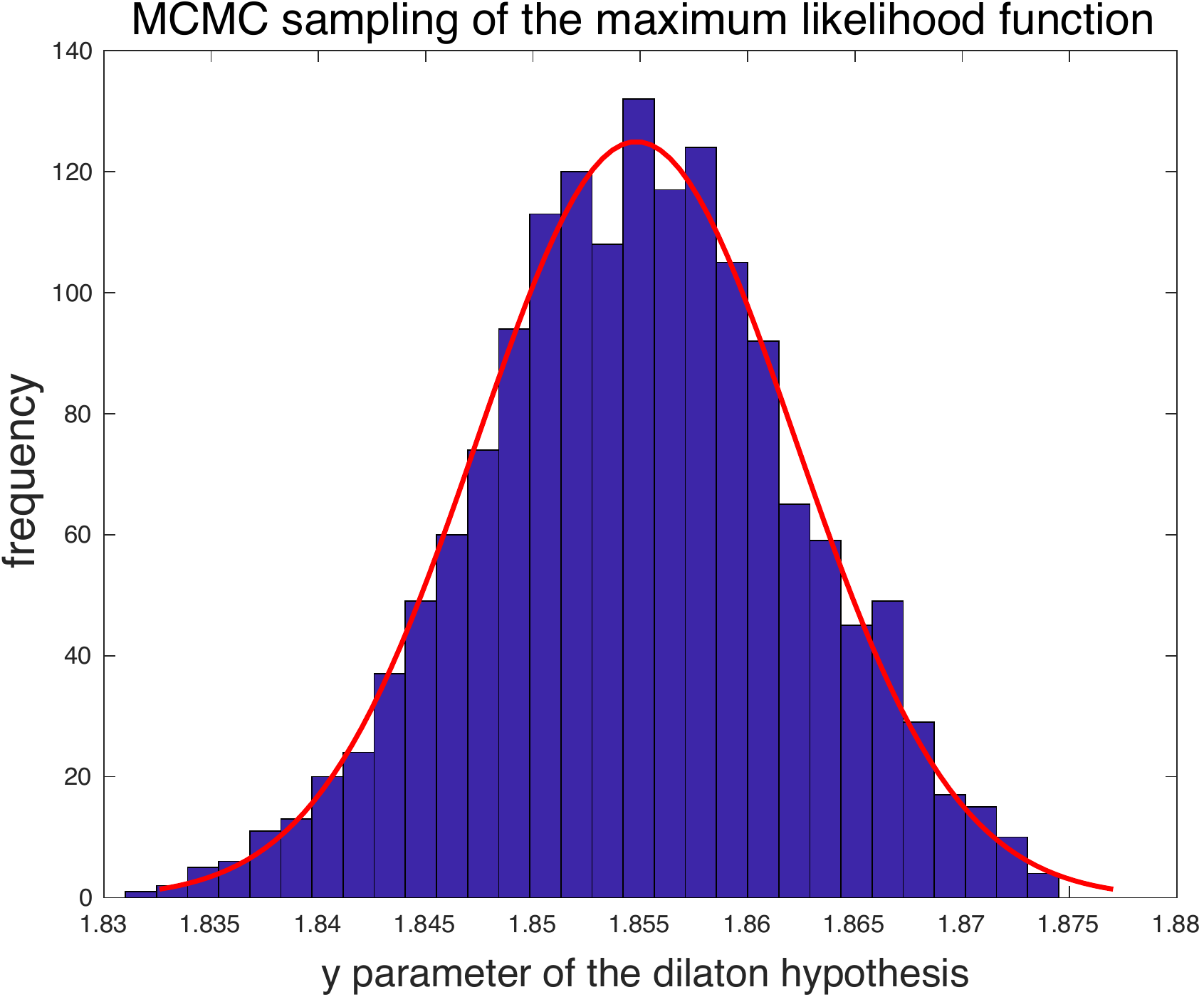}
	\includegraphics[width=2.95 in]{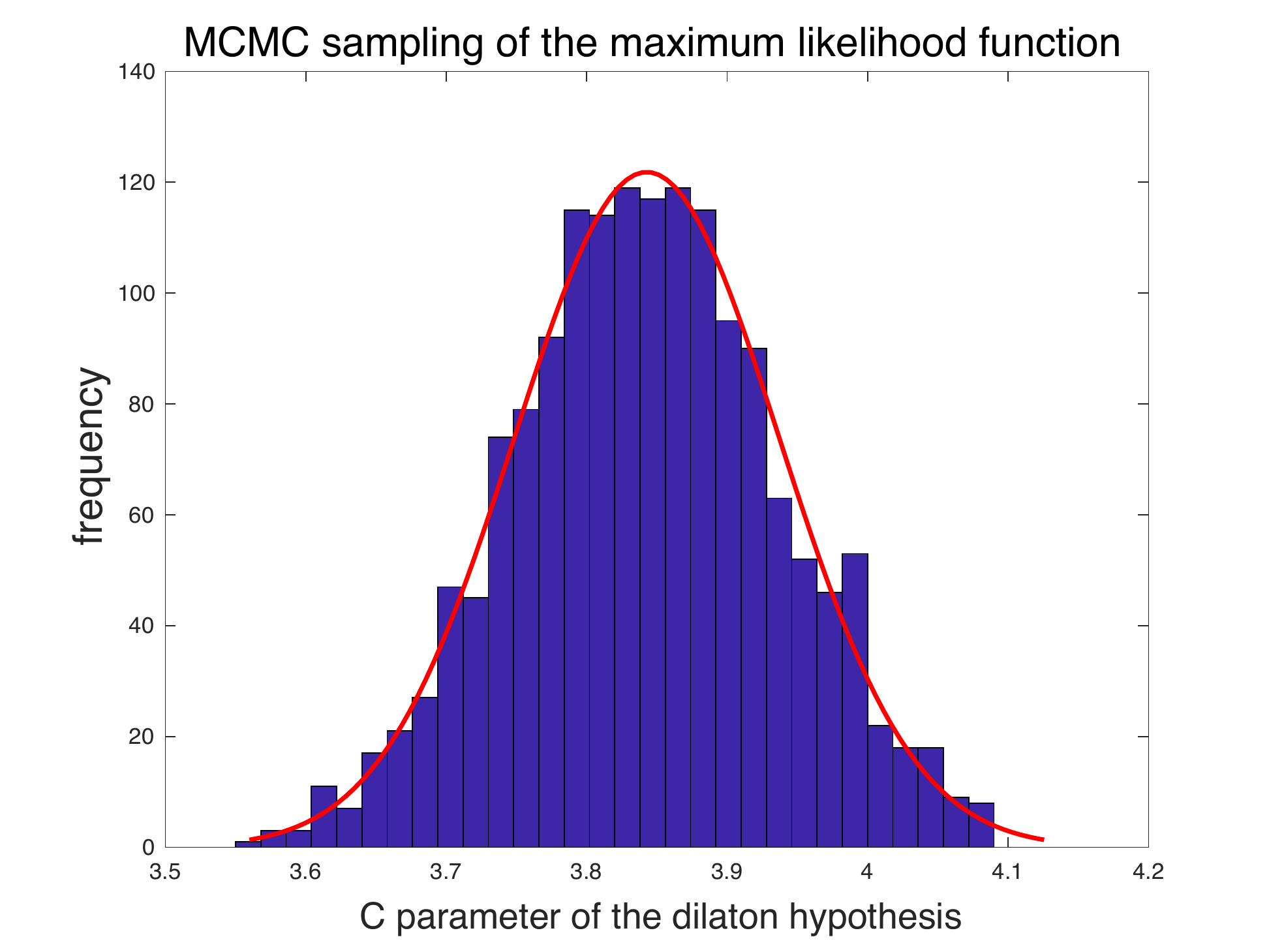}
	\caption{\label{fig:MCMC} Results are shown from MCMC sampling of the parameter space $M_\pi$, $F_\pi$, $c_M$, $c_F$
		to sample the $y$ and $C$ parameters of the dilaton scaling relation at $\beta=3.20$. 
		Each of the five MCMC runs generated $10^5$ sampled points 
		and every 50th was used in the y-distribution and C-distribution. The red curves are  fits of normal distributions to the samples. 
		These are completely independent results from the $\chi^2$ fitting procedures shown in Fig.~\ref{fig:dilaton1} and confirm their
		interpretation as Maximum Likelihood based $\chi^2$ probability distributions.}
\end{figure}
As input we use the $M_\pi$, $F_\pi$ pairs with normal 
error distributions which can be tested outside the dilaton fitting procedure. $\chi^2$ test for the
expression $M_\pi^2\cdot F_\pi^{2-y}$ would require normal error distribution for this non-linear quantity. The errors themselves 
are obtained at each $m$ from the covariance matrices of the fitted parameters $M_\pi$, $F_\pi$, $c_M$, $c_F$ from 
simultaneous fits to finite volume data $M_\pi(L)$ and $F_\pi(L)$ at each $m$.  Only the $2\times 2$ covariance matrix of the fitted 
infinite volume parameters $M_\pi$, $F_\pi$  are used in the error estimates of $M_\pi^2\cdot F_\pi^{2-y}$ at fixed lattice spacing. 
The normal distribution of the errors 
is determined from the Bayesian posterior parameter distribution of $M_\pi$, $F_\pi$, $c_M$, $c_F$ in the 4-dimensional parameter space using
Markov Chain Monte Carlo (MCMC) sampling of the parameter distributions in the Maximum Likelihood function 
at each of the five fermion masses. 
The distributions of the fitted parameters $y$ and $C$ are shown in Fig.~\ref{fig:MCMC} from 
the Bayesian posterior distribution of the $M_\pi$, $F_\pi$, $c_M$, $c_F$ parameters, MCMC sampled at five $m$ inputs and refitted for 
the $y$ and $C$ parameters. The $y$ and $C$  parameters have normal distributions and they are in perfect agreement with  $\chi^2$ fits in
Fig.~\ref{fig:dilaton1}.

\begin{figure}[htb!]
	\centering
	\includegraphics[width=3 in]{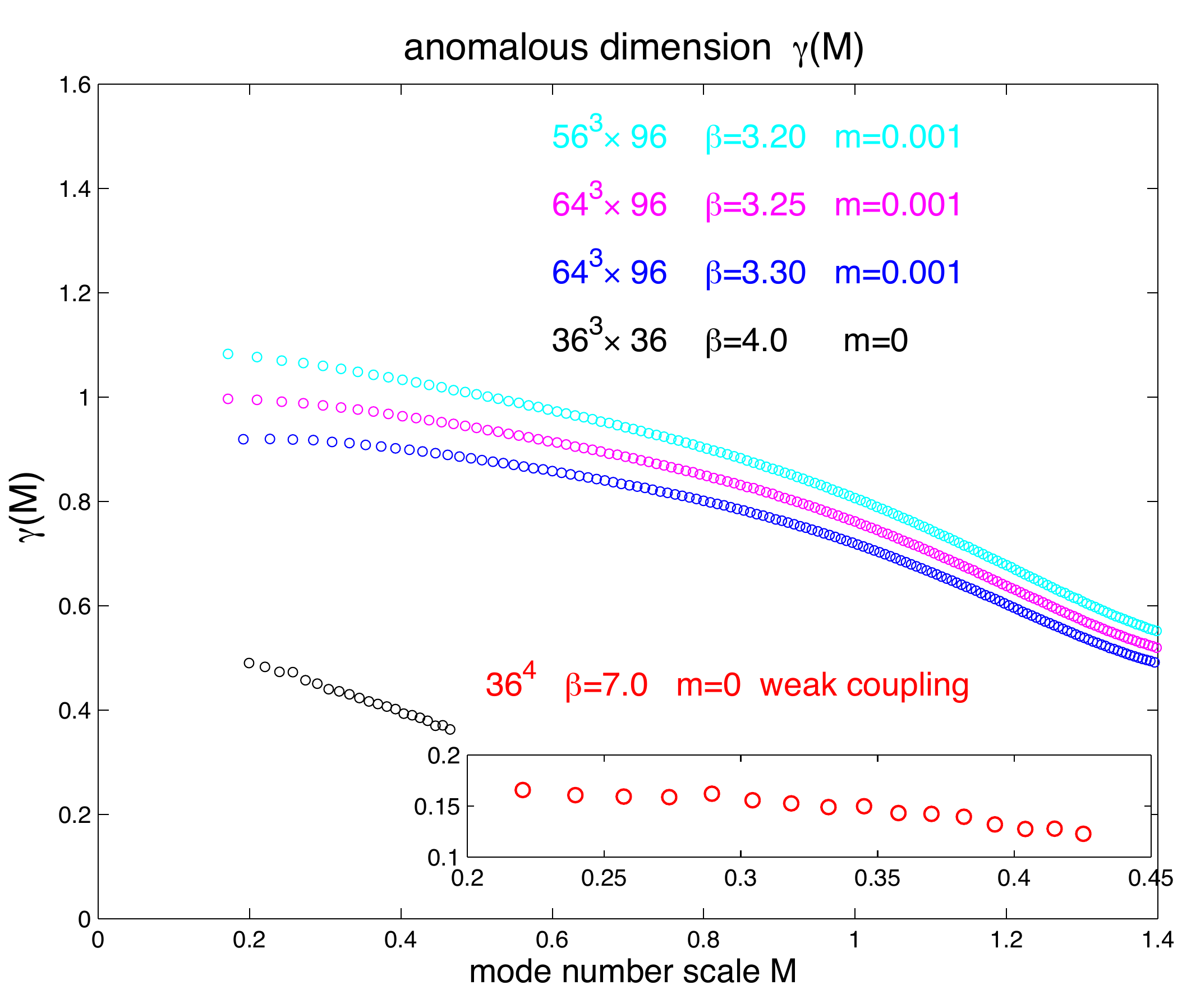}
	\caption{\label{fig:dirac}\footnotesize Results for the mass anomalous dimension $\gamma(M)$,
		are shown for the sextet model with a fermion  doublet in the two-index symmetric (sextet) representation 
		of the SU(3) color gauge group. There are results at five different lattice spacings, one of them at exactly zero 
		fermion mass. 
    	The value of $\gamma$ depends on the fermion mass, the lattice spacing and on the scale of the Dirac spectrum. }
\end{figure}

\vskip 0.1in
\noindent{\em Mass anomalous dimension:}
Using our novel Chebyshev expansion based algorithm for the mode number and spectral density of the Dirac spectrum~\cite{Fodor:2016hke} 
we developed two independent determinations of the mass anomalous dimension $\gamma$ which is playing a critically important
role in the tests of the dilaton effective theory. Typical raw data for $\gamma$ are shown in 
Fig.~\ref{fig:dirac} from the exponent of the mode number distribution at five different lattice spacings in the sextet model. 
We prefer now the more sophisticated method using the step function of the renormalization constants $Z_p$ 
which is determined from  the mode number distribution in the p-regime of $\chi SB$.
The value of $\gamma$ depends on the fermion mass, the lattice spacing and on the scale of the Dirac spectrum 
with important implications for the dilaton analysis.

Several issues  require continued ongoing work on scaling tests of the dilaton effective theory:

\begin{enumerate}[(a)]
	
\item Although values of $\gamma$ in the infrared limit are close to one, as shown in Fig.~\ref{fig:dirac}, consistent with independent fits 
of the dilaton tests, it is far from clear exactly what scale is relevant in $\gamma$ for the dilaton analysis. Fig.~\ref{fig:dilaton1} shows 
significant cutoff dependence in the fits of $\gamma$ at two different lattice spacings.  The precise meaning of 
the relevant scale set in $\gamma$ is also left undetermined. For walking theories 
the standard argument is to take $\gamma$ at the gauge coupling where the $\beta$ function is 
minimal in the sign convention we use~\cite{Golterman:2016hlz}. This is an unchecked proposition for the sextet model and we do not 
wish to comment on the $n_f=8$ model with its $\beta$-function much further away from zero and with a heavier $n_f=8$ scalar.

\item The cutoff-dependence has to be addressed in a more systematic fashion, not only for the mass anomalous dimension $\gamma$
but for all targeted physical quantities.

\item The exponent $2-y$ in $F_\pi^{2-y}$ is very close to zero. Expanding in the small exponent it is not clear if the behavior captured 
by the dilaton test is not a shadow effect of chiral perturbation theory.

\item In testing the asymptotic form of the dilaton potential $V(\chi)$ we ran into problems with the fit results whose origin remains unclear as 
we will discuss next.

\end{enumerate}

\subsection{Tests of the asymptotic dilaton potential  ${\rm \boldsymbol{V(\chi) \sim \chi^p}}$}

\begin{figure}[htb!]
	\centering
	\includegraphics[width=0.49\linewidth]{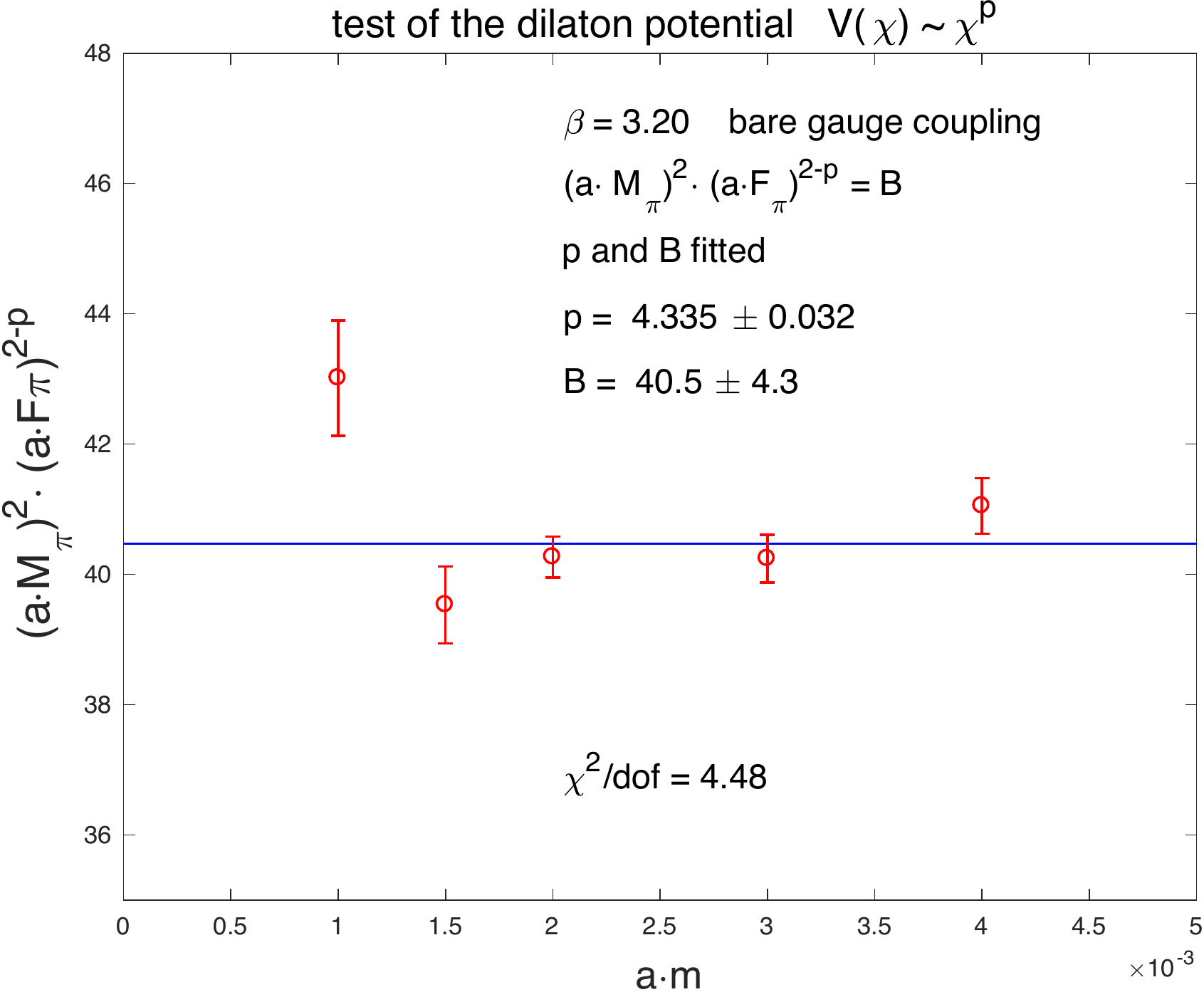}
	\includegraphics[width=0.49\linewidth]{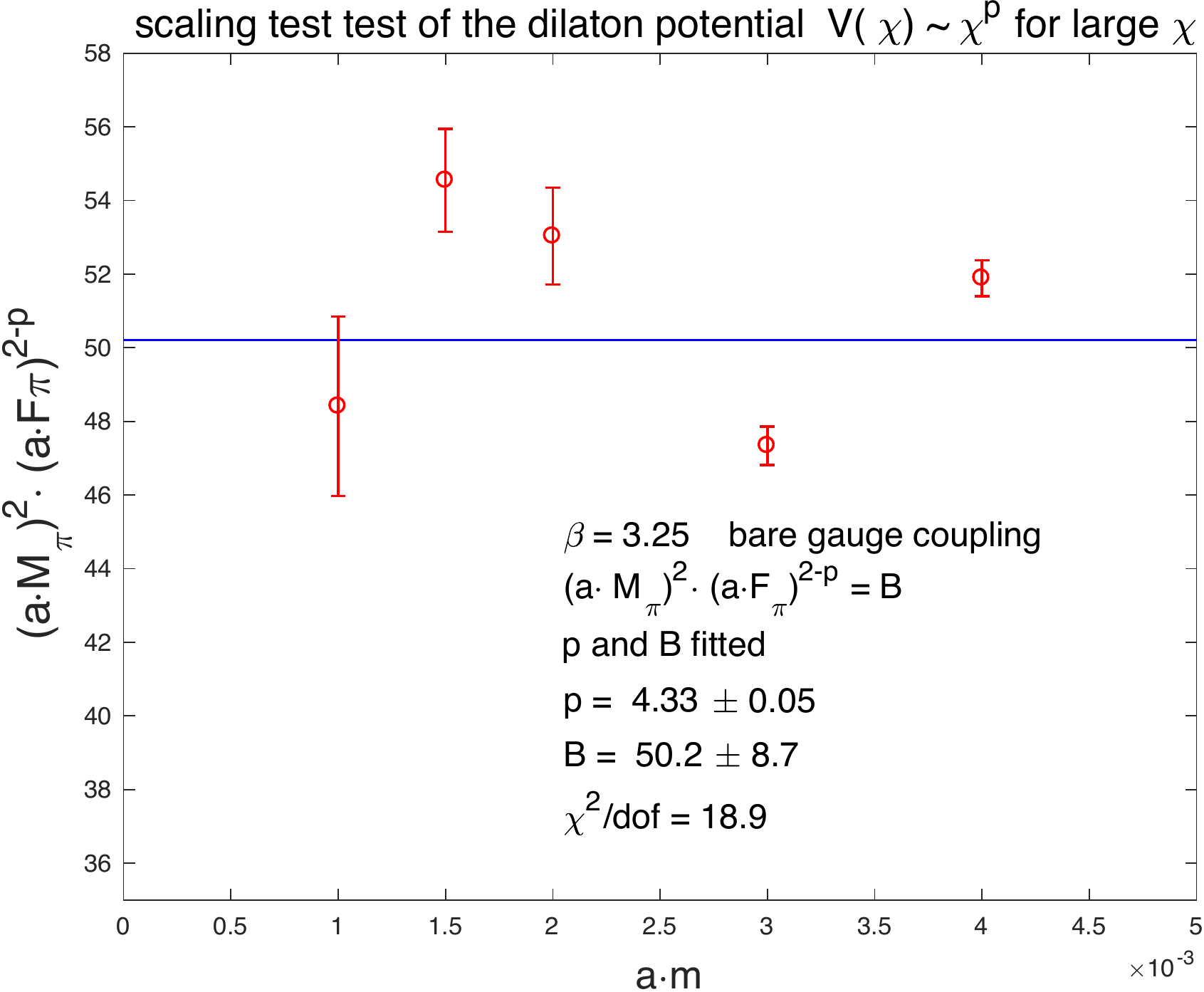}
	\caption{\label{fig:dilaton2} The FSS corrected values of $M_\pi$ and $F_\pi$ are used as input for the 
		statistical analysis of the dilaton test for asymptotic behavior of the dilaton potential.}
\end{figure}

A tree-level scaling test was proposed in~\cite{Appelquist:2017wcg} which is dependent on the asymptotic form of
the dilaton potential. The results for the sextet model are shown in Fig.~\ref{fig:dilaton2} for two 
gauge couplings.
Accordingly, a general fingerprint of the asymptotic dilaton potential is given by the constraint of the
relation $M_\pi^2\cdot F_\pi^{2-p} = B$ where the exponent $p$ is related to the asymptotic form of the 
dilaton potential $V(\chi) \sim \chi^p$ for large $\chi$. The constant $B$ is related to the amplitude of the leading term in the dilaton
potential. The test essentially amounts to showing that $M_\pi^2\cdot F_\pi^{2-p} $ is independent from the fermion mass $m$
and the constant value sets $B$.

The FSS corrected values of $M_\pi$ and $F_\pi$ are used again as input data for the statistical analysis with results shown in the figure.
The statistical analysis of the fitting procedure is very similar to what was used above in the potential-independent 
scaling tests of the effective dilaton theory.
As input we use the $M_\pi$, $F_\pi$ pairs with normal 
error distributions which can be tested outside the dilaton fitting procedure. $\chi^2$ test for the
expression $M_\pi^2\cdot F_\pi^{2-p}$ would require normal error distribution for this non-linear quantity. The errors themselves 
are obtained at each $m$ from the covariance matrices of the fitted parameters $M_\pi$, $F_\pi$, $c_M$, $c_F$ from 
simultaneous fits to finite volume data $M_\pi(L)$ and $F_\pi(L)$ at each $m$.  Only the $2\times 2$ covariance matrix of the fitted 
infinite volume parameters $M_\pi$, $F_\pi$  are used in the error estimates of $M_\pi^2\cdot F_\pi^{2-p}$ at fixed lattice spacing. 
The normal distribution of the errors 
is determined from the Bayesian posterior parameter distribution of $M_\pi$, $F_\pi$, $c_M$, $c_F$ in the 4-dimensional parameter space using
Markov Chain Monte Carlo (MCMC) sampling of the parameter distributions in the Maximum Likelihood function 
at each of the five fermion masses. The two procedures are consistent with each other.

The outcome of the test results for the leading term of the dilaton potential remains controversial. 
The $\beta=3.20$ test has one outlier at the lowest fermion mass but the $\beta=3.25$ test is not acceptable in its current form. 
While looking for explanations, so far without results, it is natural to raise the question if the quality of the data set can really differentiate
between dilaton and $\sigma$-model scenarios. The more robust general test yielded an exponent $y$
close to $y=2$ which perhaps could be accommodated in 
some generalized effective action of the $\sigma$-model with added terms and with loop corrections~\cite{Soto:2011ap,Hansen:2016fri}.  
This is illustrated next with the conventional  fitting procedure in chiral perturbation theory.

\section{Chiral perturbation theory strikes back}

\begin{figure}[htb!]
	\centering
	\includegraphics[width=0.49\linewidth]{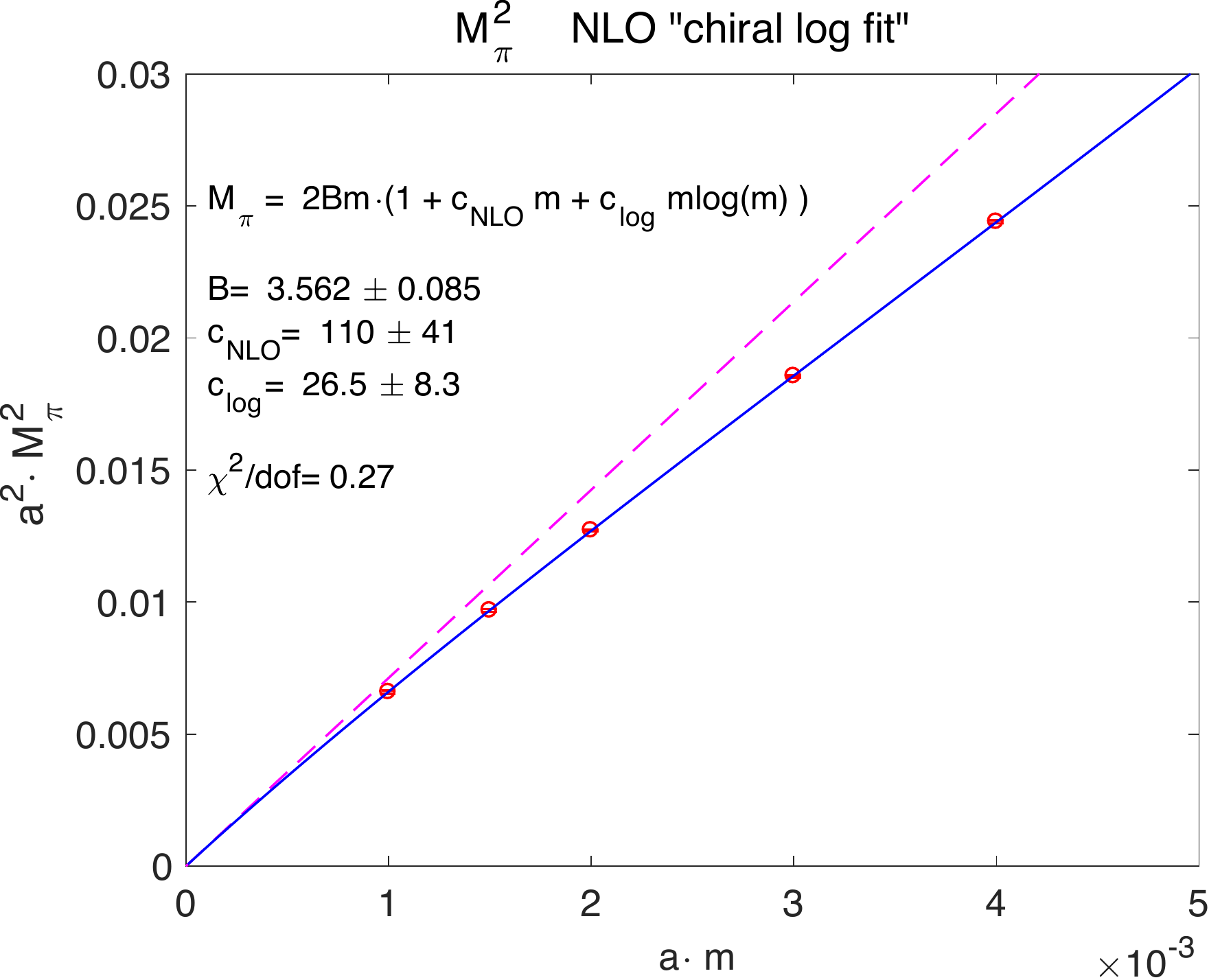}
	\includegraphics[width=0.48\linewidth]{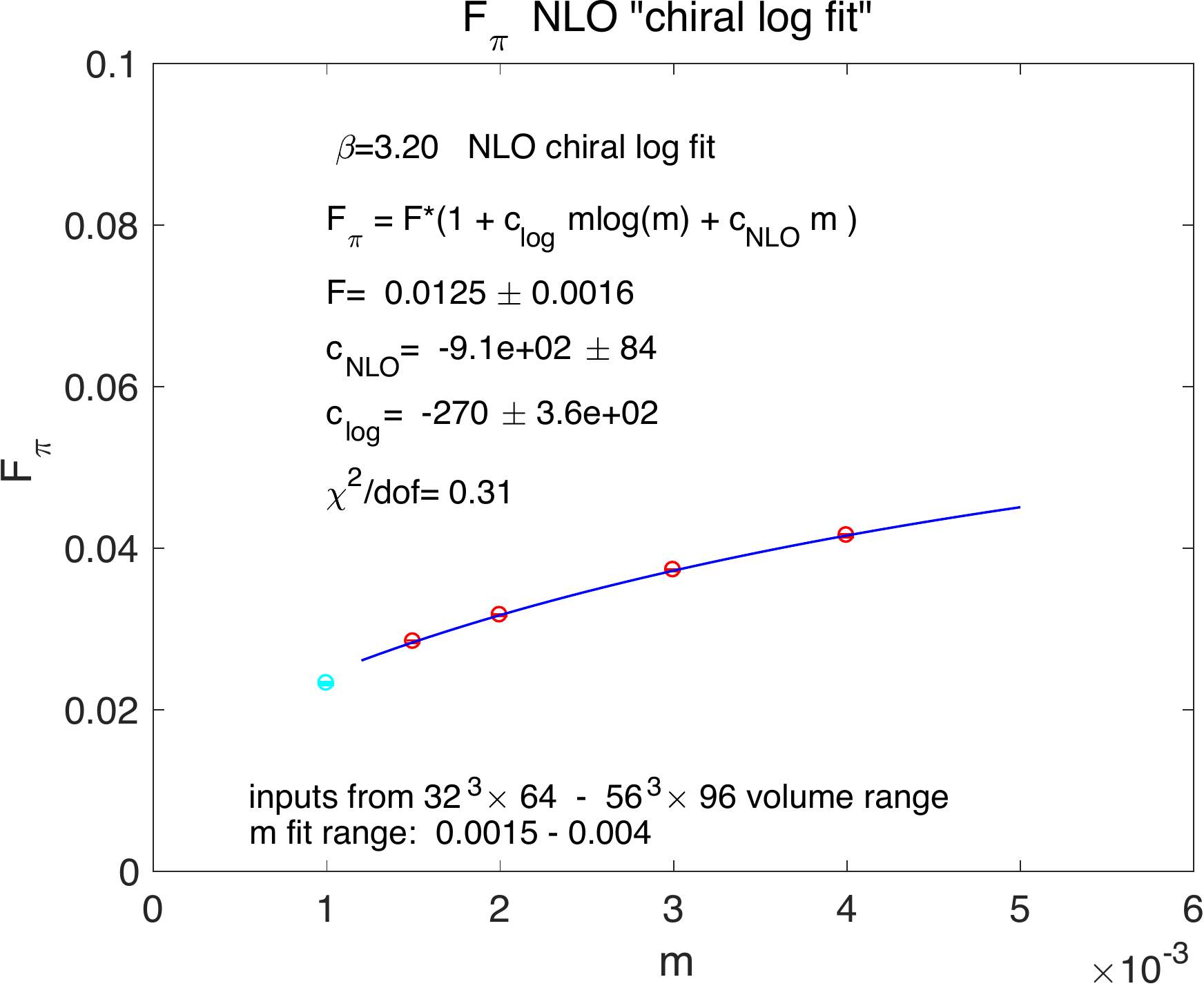}
	\caption{\label{fig:chiralPT} One-loop continuum chiral fits are shown for $M_\pi$ and  $F_\pi$ at $\beta=3.20$.}
\end{figure}

Our conventional chiral fits for $M_\pi$ and $F_\pi$ are shown with logarithmic NLO loop corrections in Fig.~\ref{fig:chiralPT}.
The most recent FSS corrected $M_\pi$ and $F_\pi$  input data  are used, the same what was used in the dilaton fits. 
The "NLO fits" are good although there is an outlier in $F_\pi$ at the lowest $m$ value, like the outlier in the dilaton potential fit
at $\beta=3.20$. 

The fits ignore taste breaking and other cutoff effects which lead to the inconsistent determination of 
the fundamental parameters $B$ and $F$ of the chiral Lagrangian. 
It was shown earlier~\cite{Fodor:2016pls} that rooted staggered perturbation theory can be fitted in the sextet model with a consistent pair of the 
fundamental parameters $B$ and $F$. Here we show the updated fits based on the most recent FSS input as described above.
For the SU(2) analysis we adapted the procedure from~\cite{Aubin:2003mg}.
There are two fundamental parameters $F$  and $B$  in the SU(2) chiral Lagrangian
in conventional notation from~\cite{Appelquist:2017wcg}. The fundamental parameter $F$
of ${\chi PT}$, defined as the chiral limit of
the pion decay constant ${F_\pi}$, sets the Electroweak scale and the fundamental parameter $ B$ sets the 
fermion mass deformation of the Goldstone spectrum. With bare fermion mass m, the 
RG invariant combination ${ m\cdot B\cdot F^2}$ is related 
to the chiral condensate via the GMOR relation. 

We apply rooted staggered chiral perturbation theory to the mass-deformed pion spectrum and $F_\pi$. 
The fitting procedure in the p-regime proceeds in several steps. 

\begin{figure}[htb!]
	\begin{center}
		\begin{tabular}{cc}
			\includegraphics[height=5cm]{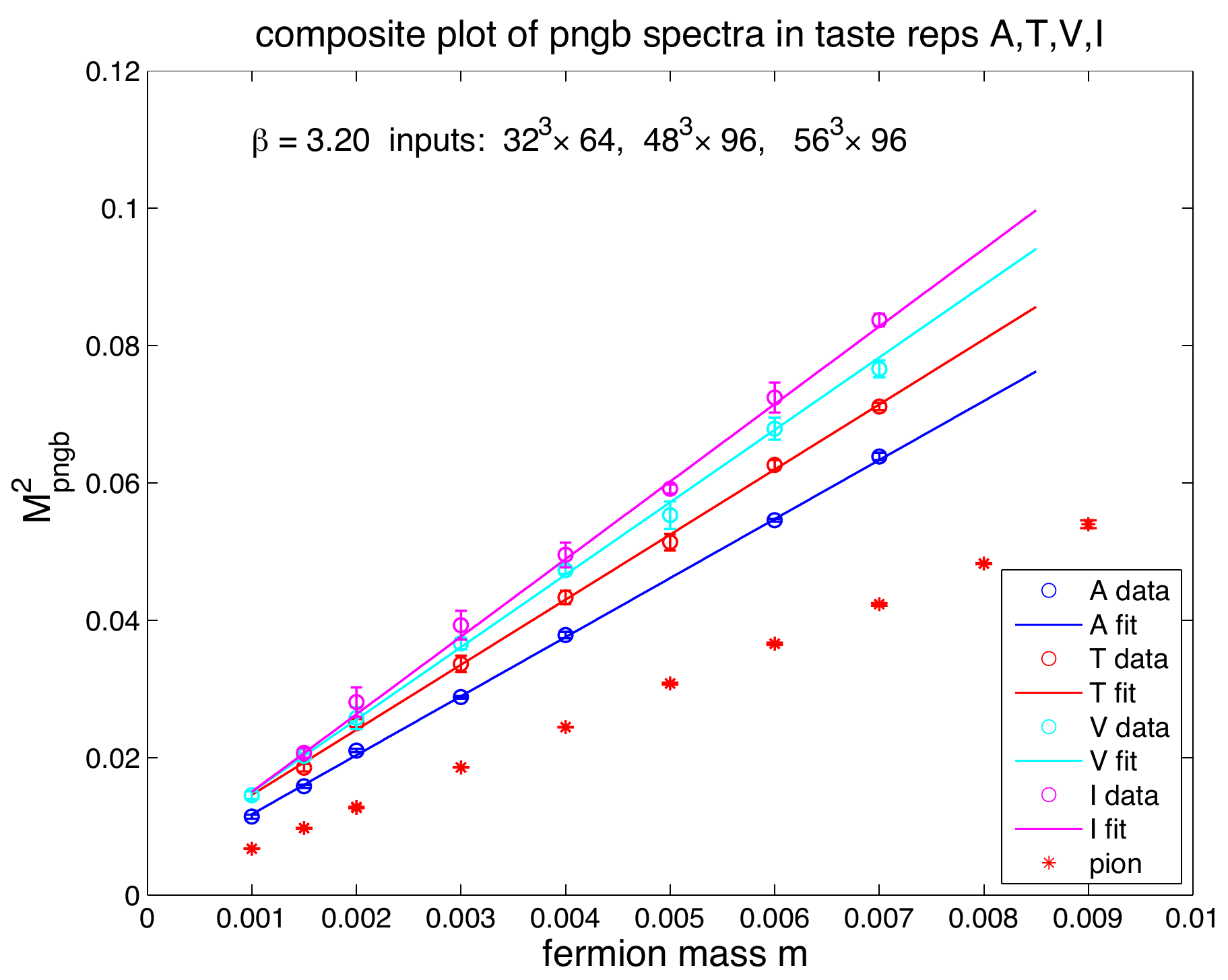}&
			\includegraphics[height=5cm]{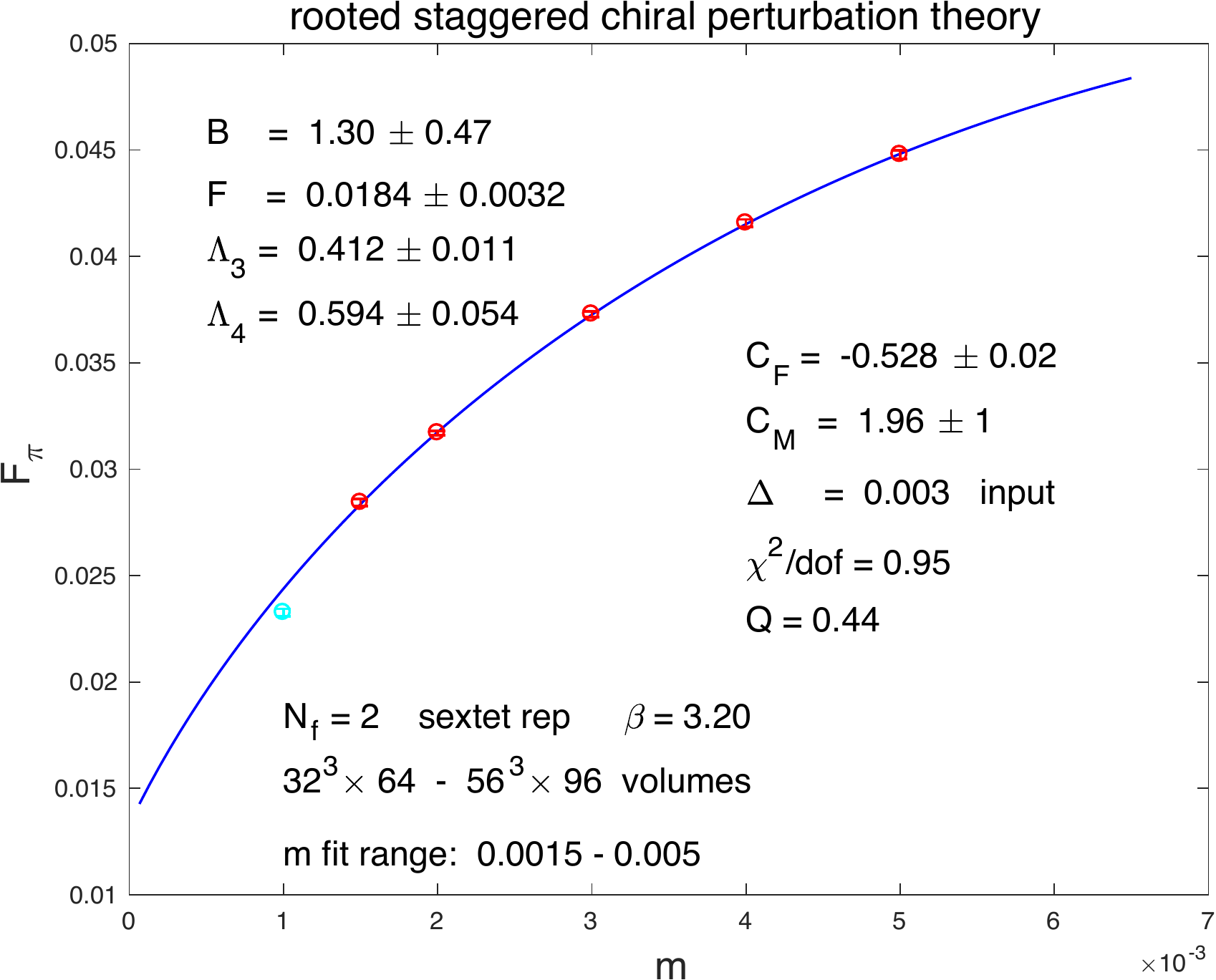}\\
			\includegraphics[height=5cm]{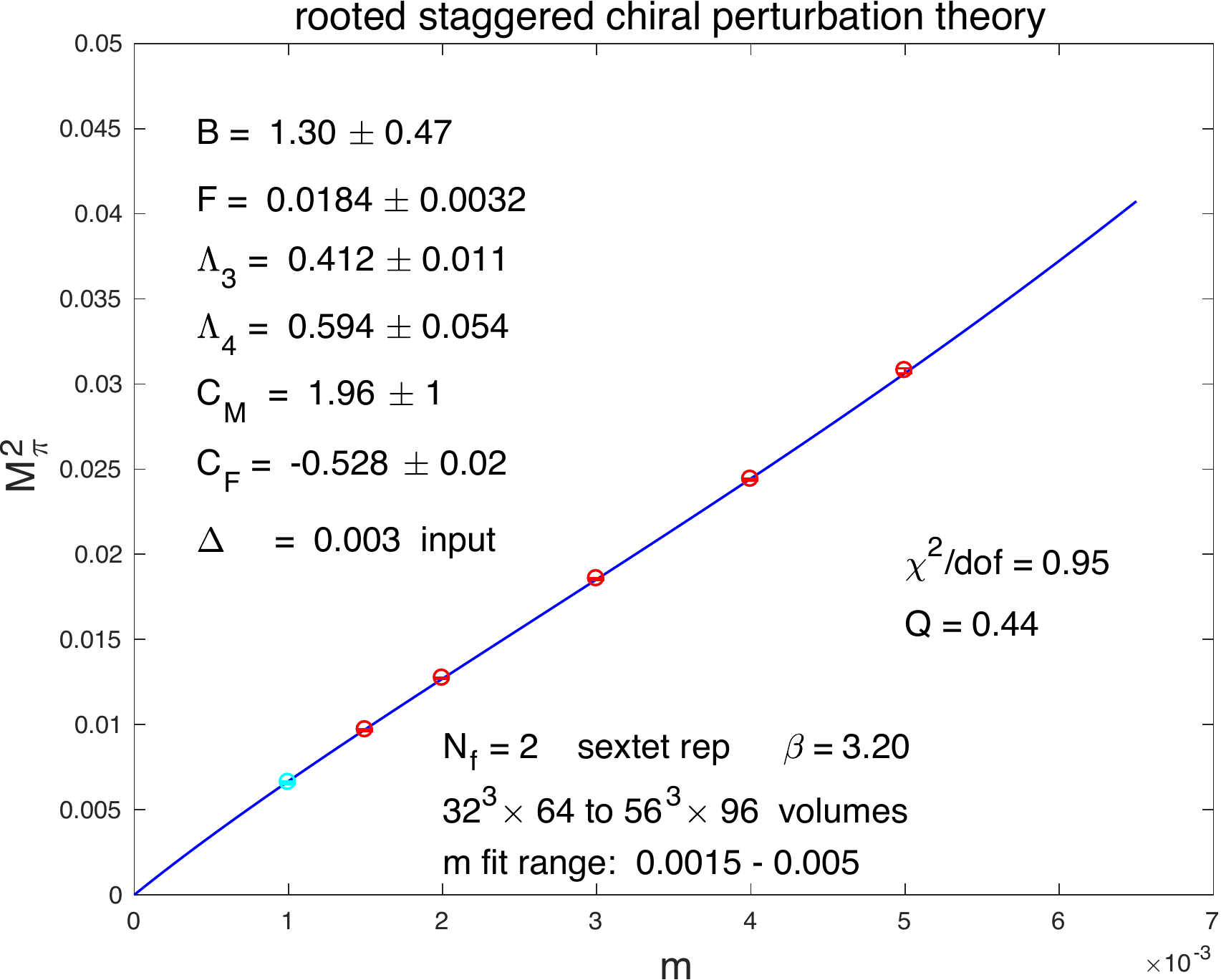}&
			\includegraphics[height=5cm]{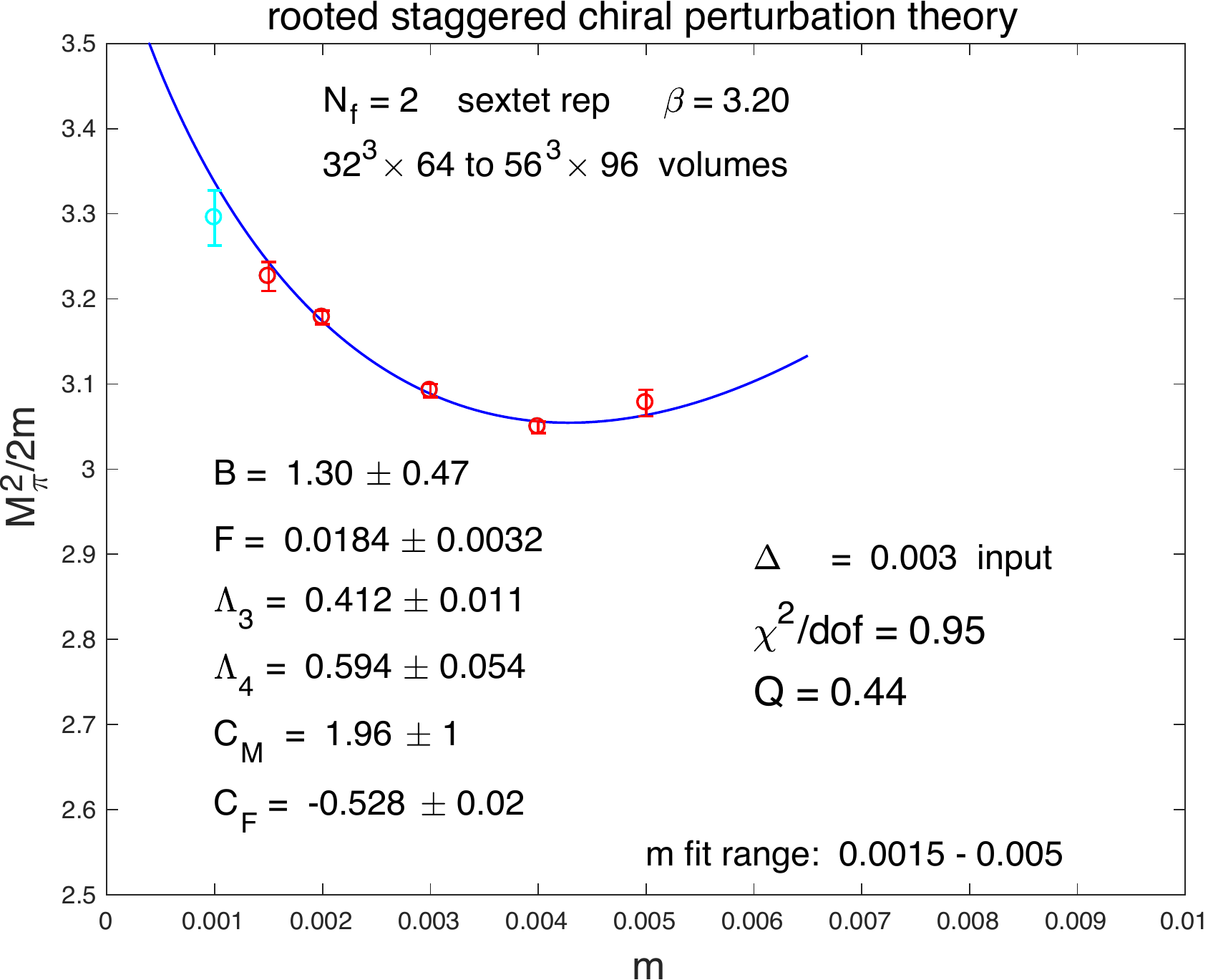}
		\end{tabular}
	\end{center}
	\caption{\footnotesize  Results from rooted ${ \chi PT}$ are shown
		from fits at gauge coupling ${ \beta = 3.20}$ 
		which corresponds to our coarser lattice of the two extended sets of gauge ensembles. 
		The upper left panel shows the linear fits to the quadratic masses of the non-Goldstone pions to determine their 
		mass shifts and slopes as input. The upper right panel shows the 
		rooted ${\chi PT}$ fit to ${ F_\pi}$ as a function of fermion mass
		deformations away from the chiral limit. The two lower panels  show  
		rooted ${\chi PT}$ fits to ${M_\pi}$ as a function of fermion mass
		deformations away from the chiral limit. 
		We have similar analysis for ${M^2_{\pi}}$ and ${ F_{\pi}}$ at ${ \beta=3.25}$. }
	\label{fig:chipt}
\end{figure}

\begin{enumerate}[(a)]

\item In the first step  
finite volume correction is applied to the ${ M_\pi}$ and ${ F_\pi}$ data with 1-loop continuum ${ \chi PT}$ inspired Ansatz.
This FSS procedure was described above. 

\item
A linear fit is applied to the quadratic masses of the non-Goldstone pion spectrum to determine their 
mass shifts and slopes.

\item
In the final analysis of rooted chiral perturbation theory, non-Goldstone pion states run in the chiral loops 
including their mass splittings and fan-out slope structure from taste breaking as determined from the  
linear fits to the non-Goldstone spectrum. 
We applied this analysis at two values of the gauge coupling where we have extensive ensembles as shown in Fig.~\ref{fig:chipt}
for $\beta=3.20$.

\end{enumerate}

The fitting procedure in Fig.~\ref{fig:chipt} only serves as a feasibility study for illustration. 
Several "successful" fits lead to consistent $B,F$ pairs.
The results from ${ rs\chi PT}$ are shown in Figure~\ref{fig:chipt}
from fits at gauge coupling ${ \beta = 3.20}$ 
which corresponds to our coarser lattice of the two extended sets of gauge ensembles used in the studies of this conference submission. 
The upper left panel shows the linear fits to the quadratic masses of the non-Goldstone pions to determine their 
mass shifts and slopes as input. The upper right panel shows the 
${ rs\chi PT}$ fit to ${ F_\pi}$ as a function of fermion mass
deformations away from the chiral limit. The two lower panels  show  
${ rs\chi PT}$ fits to ${ M_\pi}$ as a function of fermion mass
deformations away from the chiral limit. 
Fits at the finer lattice 
spacing ${ \beta = 3.25}$ are quite similar in quality but with lower confidence level. 
The unambiguous determination of the cutoff dependent $F$  and $B$ parameters and their continuum limit 
from  ${ rs\chi PT}$ would require extended analysis.

Although our results are consistent with chiral symmetry breaking and ${ rs\chi PT}$,  continued work would
require considerable  extensions for definitive results. Most importantly, a solution to the entanglement problem of the
light $\sigma$-particle with  low-energy pion dynamics would require new analysis based on some extended effective theory
of the $\sigma$-model as an alternative to the attractive dilaton scenario.

\section{Conclusions and outlook}
Based on our most recent analysis, we have presented some successes and some puzzles when an effective dilaton theory is applied to the sextet model.
Perhaps the consistent description of the light scalar as a dilaton in entanglement with pion dynamics 
is the most attractive feature of this framework. Chiral perturbation theory is not ruled out by the data. In fact consistent fits 
can be found but they are ambiguous in choosing parameter sets and most importantly would require some extended effective description
like the extended $\sigma$-model as an alternative to the attractive dilaton scenario. 
This is a rapidly developing and exciting research area with two interesting frameworks and with important BSM implications.

\section*{Acknowledgments}
{
	We acknowledge support by the DOE under grant DE-SC0009919, by the NSF under grants 1318220 and 1620845, 
	by OTKA under the grant OTKA-NF-104034, and by the Deutsche
	Forschungsgemeinschaft grant SFB-TR 55. Computational resources were provided by the DOE INCITE program
	on the ALCF BG/Q platform, USQCD at Fermilab, 
	by the University of Wuppertal, by Juelich Supercomputing Center on Juqueen
	and by the Institute for Theoretical Physics, Eotvos University. We are grateful to Szabolcs Borsanyi for his code 
	development for the BG/Q platform. We are also 
	grateful to Sandor Katz and Kalman Szabo for their CUDA code development.
}

\bibliography{jkNf12}

\end{document}